\documentclass[aps,twocolumn,superscriptaddress,floatfix,longbibliography]{revtex4-2}
\usepackage{amsmath,amssymb,amsthm}
\usepackage{physics}
\usepackage{amsfonts}
\usepackage{mathrsfs}
\usepackage{graphicx}
\usepackage{tabularx}
\usepackage{enumerate}
\usepackage{dcolumn}
\usepackage{bm}
\usepackage{xcolor}
\usepackage[normalem]{ulem}
\usepackage[colorlinks,linkcolor=blue,citecolor=blue,urlcolor=blue]{hyperref}

\begin{document}
\title{Exact anomalous mobility edges in one-dimensional non-Hermitian quasicrystals}
\author{Xiang-Ping Jiang}
\affiliation{Zhejiang Lab, Hangzhou 311121, China}

\author{Weilei Zeng}
\address{Zhejiang Lab, Hangzhou 311121, China}

\author{Yayun Hu}
\email{yyhu@zhejianglab.edu.cn}
\affiliation{Zhejiang Lab, Hangzhou 311121, China}

\author{Lei Pan}
\email{panlei@nankai.edu.cn}
\affiliation{School of Physics, Nankai University, Tianjin 300071, China}


\date{\today}

\begin{abstract}
Recent research has made significant progress in understanding localization transitions and mobility edges (MEs) that separate extended and localized states in non-Hermitian (NH) quasicrystals. Here we focus on studying critical states and anomalous MEs, which identify the boundaries between critical and localized states within two distinct NH quasiperiodic models. Specifically, the first model is a quasiperiodic mosaic lattice with both nonreciprocal hopping term and on-site potential. In contrast, the second model features an unbounded quasiperiodic on-site potential and nonreciprocal hopping. Using Avila's global theory, we analytically derive the Lyapunov exponent and exact anomalous MEs. To confirm the emergence of the robust critical states in both models, we conduct a numerical multifractal analysis of the wave functions and spectrum analysis of level spacing. Furthermore, we investigate the transition between real and complex spectra and the topological origins of the anomalous MEs. Our results may shed light on exploring the critical states and anomalous MEs in NH quasiperiodic systems.
\end{abstract}

\maketitle

\section{Introduction}
Anderson localization is a fundamental phenomenon in which quantum wavefunctions become exponentially localized in the presence of random disorder, without the tendency to diffuse~\cite{anderson1958absence}. In one and two-dimensional quenched disorder systems, one-parameter scaling theory predicts that all noninteracting eigenstates become localized even with arbitrarily infinitesimal disorder strength\cite{thouless1974electrons,abrahams1979scaling,evers2008anderson}. However, in three-dimensional disorder systems, it has been demonstrated that localized and extended states can coexist at finite levels of disorder, with a critical energy known as the mobility edge (ME) acting as a boundary between these two phases. Comparatively, one-dimensional (1D) quasiperiodic systems can exhibit unique behaviors and undergo localization transitions. A prototypical example is the Aubry-Andr{\'e}-Harper (AAH) model~\cite{harper1955single,aubry1980analyticity}, which undergoes a localization transition when the strength of the quasiperiodic potential exceeds a critical threshold. The AAH model is renowned for its exact solvability, offering significant benefits for obtaining exact results due to the self-duality between real and momentum spaces~\cite{gonccalves2022hidden,borgnia2022rational,borgnia2023localization,gonccalves2023critical,gonccalves2023renormalization,vu2023generic,lin2023general}. Through the investigation of various extensions of the AAH models, experimental and theoretical researchers have discovered evidence for the existence of energy-dependent MEs in 1D generalized AAH models~\cite{sarma1988mobility,boers2007mobility,biddle2009localization,modugno2009exponential,biddle2010pre,ganeshan2015nearest,danieli2015flat,luschen2018single,liu2018mobility,wang2020one,wang2023two,wang2023engineering,qi2023multiple,li2023anderson}.

In 1D quasiperiodic systems, three primary quantum states have been observed: extended, localized, and critical states. Critical states are extended yet non-ergodic, showing local scale invariance and possessing fundamentally distinct properties in terms of spectral statistics, multifractal characteristics, and dynamical evolution compared to localized and extended states. Conventionally, MEs have been employed to distinguish between localized and extended states. However, recent advances in research have introduced a novel type of MEs referred to as anomalous mobility edges (AMEs)~\cite{liu2022anomalous,zhang2022lyapunov,zhang2022anderson,zhou2023exact,liu2023predict,lee2023critical1}, which serve as boundaries between critical states and localized states. These discoveries and analyses of AMEs have significantly advanced our comprehension of critical states and the localization phenomena in quasiperiodic systems~\cite{ahmed2022flat,lee2023critical}. 

In recent years, there has been an escalating interest in the examination of Anderson localization and MEs in non-Hermitian (NH) disordered and quasiperiodic systems~\cite{zeng2017anderson,longhi2019topological,longhi2019metal,jiang2019interplay,zeng2020topological1,zeng2020topological2,zhai2021cascade,jiang2021mobility,jiang2021non,liu2021exactb,weidemann2022topological,schiffer2021anderson,chen2022breakdown,zhu2023topological,gandhi2023topological,liu2022real,acharya2022localization,zhou2023non,ghosh2023eigenvector,acharya2024localization,zhou2024entanglement1,li2024emergent,jiang2024localization}. Typically, NH systems are constructed by incorporating nonreciprocal hopping processes or gain and loss terms into their Hamiltonians. For example, with the NH extensions of the AAH model through the complexification of the potential phase, it has been demonstrated that the localization transitions exhibit a topological nature and are characterized by winding numbers of the energy spectrum. Meanwhile, the concept of the ME has also been extended to NH systems. It has been found that the ME can be used to predict the boundary of extended states and the transition from real to complex energy spectrum for NH quasiperiodic systems, thereby introducing a topological signature of MEs~\cite{liu2020non,liu2020generalized,zeng2020winding,longhi2021non1,liu2021exacta,longhi2022non1,mu2022non,xu2022exact,xia2022exact,qi2023localization,wang2024exact,jiang2024exact,wang2024non,li2024ring}.  Despite extensive studies on the effects of non-Hermiticity on localization transitions and traditional MEs in various contexts, investigation of critical states and AMEs alongside localization transitions in NH quasiperiodic models remains lacking. It remains unclear whether critical states and AMEs exist stably in NH quasiperiodic lattices. If so, how do we characterize the AMEs and whether any correlation exists between the critical-localized state transitions and the real-complex spectrum transition?

In this work, we introduce two distinct nonreciprocal NH quasiperiodic models to address the issues above. We endeavor to investigate robust critical states and exact AMEs by employing Avila's global theory, which accurately characterizes critical regions and AMEs. By analyzing the spatial distribution of wave functions and level spacings of the eigenvalues, we discover that an increase in quasiperiodic potential strength results in a critical-localized transition. This localization transition co-occurs with the real-complex spectrum transition, indicating that a winding number can describe this topological transition. Consequently, the emergence of AMEs separating critical and localized states in our models is indeed topological.

The structure of this paper is as follows. In Sec.~\ref{sec:2}, we provide a streamlined introduction to the two NH quasiperiodic models. In Sec.~\ref{sec:3}, we determine the AMEs of model I using Avila's global theory and investigate the mechanism that generates the existence of critical states. In Sec.~\ref{sec:4}, we determine the AMEs of model II. In Sec.~\ref{sec:5}, we show the real-complex spectrum transition and the topological origin of AMEs. We make a summary in Sec.~\ref{sec:6}.

\section{The model Hamiltonian}\label{sec:2}
\begin{figure}[t]
	\centering
	\includegraphics[width=0.5\textwidth]{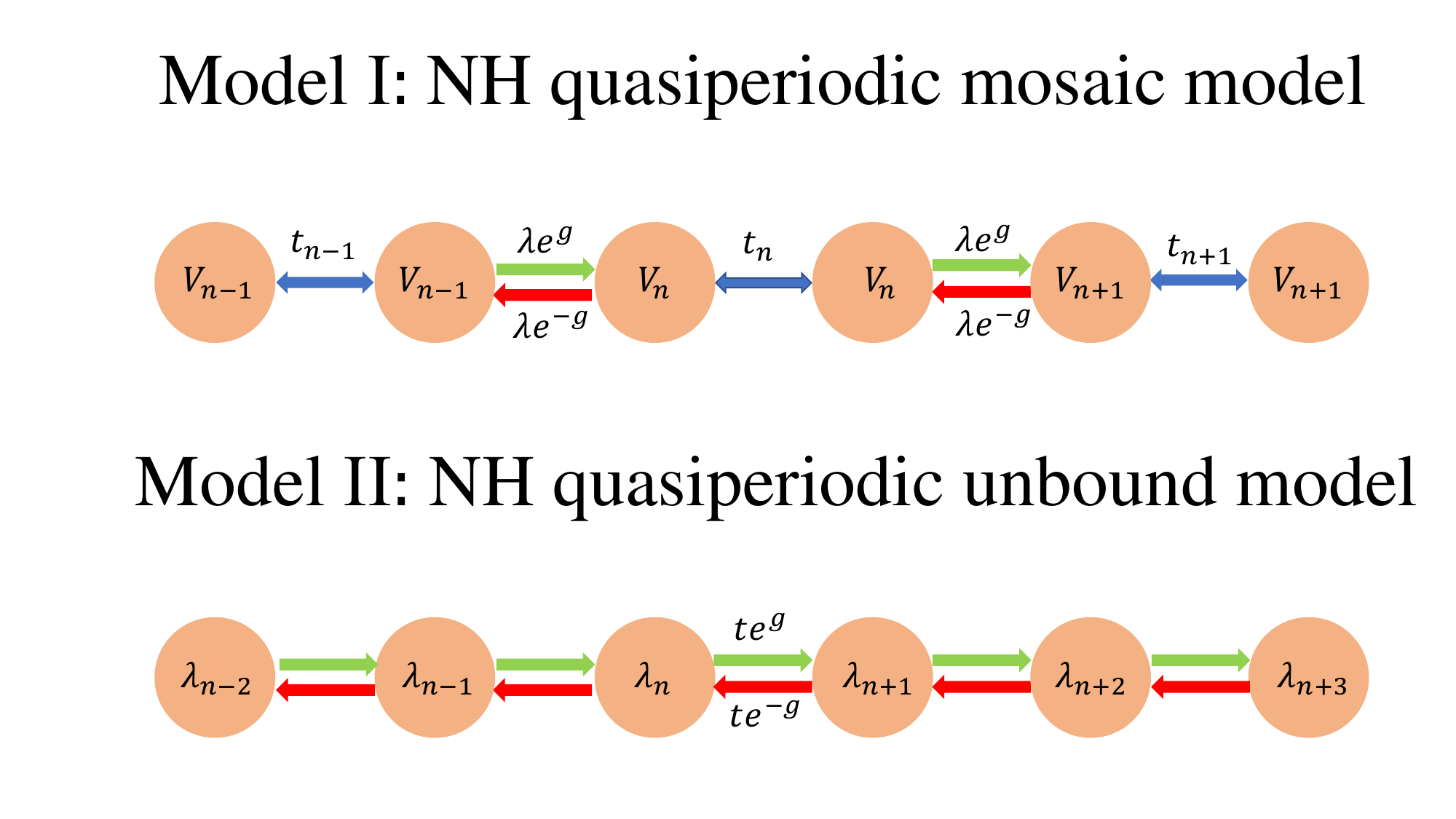}
	\caption{Schematic diagram of the models. (a) and (b) show the NH quasiperiodic mosaic model and the NH quasiperiodic unbound model, respectively. The red and green solid lines denote the nonreciprocal hopping.}
	\label{fig1}
\end{figure}

We introduce two NH quasiperiodic models that will be adopted to investigate critical states and AMEs in this work. These two models are pictorially shown in Fig.~\ref{fig1}. The Hamiltonian of model I [Fig.~\ref{fig1}(a)] is described by
\begin{equation}\label{eq:model1}
H_{\rm{I}}=\sum_{n}(t_{n}^{+}a_{n}^{\dagger}a_{n+1}+t_{n}^{-}a_{n+1}^{\dagger}a_{n})+\sum_{n}V_{n}a_{n}^{\dagger}a_{n},
\end{equation}
where $a_{n}^{\dagger} (a_{n})$ corresponds to the spinless fermion creation (annihilation) operator at site $n$. In Eq. (\ref{eq:model1}),  the critical components involve the hopping parameter  $t_{n}$ and the on-site potential $V_{n}$, both of which exhibit quasiperiodic and mosaic characteristics. The hopping coefficient $t_{n}$ is defined as
\begin{equation}\label{eq:mosaic_t1}
    t_{n}^{\pm}=\left\{\begin{matrix} 	\lambda e^{\pm g},&n=1, \mod2,\\ 	2V\cos(2\pi \alpha n+\phi),&n=0, \mod2. \end{matrix}\right.
\end{equation}
 and the on-site potential $V_{n}$ is considered as
\begin{equation}\label{eq:mosaic_v1}
    V_n=\left\{\begin{matrix} 	2V\cos[2\pi\alpha(n-1)+\phi],&n=1, \mod2,\\ 	2V\cos(2\pi \alpha n+\phi),&n=0, \mod2. \end{matrix}\right.
\end{equation}
Here $\lambda$, $g$, and $\phi$ denote the hopping coefficient, nonreciprocal strength, and phase offset. For convenience, we set on-site potential amplitude $V=1$ as unit energy.

The Hamiltonian of model II, as shown in Fig.~\ref{fig1}(b), can be written as
\begin{equation}\label{eq:model2}
H_{\rm{II}}=t\sum_{n}(e^{g}a_{n}^{\dagger}a_{n+1}+e^{-g}a_{n+1}^{\dagger}a_{n})+\sum_{n}\lambda_{n}a_{n}^{\dagger}a_{n},
\end{equation}
where $t=1$ is the hopping strength and  $\lambda_n$ is the quasiperiodic potential, which is given by    
\begin{equation}\label{eq5}
    \lambda_n=\frac{2\lambda \cos(2\pi\alpha n+\phi)}{1-b\cos(2\pi\alpha n+\phi)}.
\end{equation}
Here $\lambda$, $g$, and $b$ represent the strength of the on-site potential, the nonreciprocal strength, and the control parameter, respectively. When the NH parameter $g=0$, the Hamiltonian (\ref{eq:model2}) reduced to the Ganeshan-Pixley-Das Sarma (GPD) model~\cite{ganeshan2015nearest}, which can host the energy-dependent MEs for $\abs{b}< 1$ and AMEs for $\abs{b}\geqslant 1$~\cite{zhang2022lyapunov}. The current study examines the model's critical states and AMEs where the parameter $g\neq 0$ and $\abs{b}\geqslant 1$.

In this work, for convenience and without affecting generality, we take $\phi=0$ and $\alpha=\lim_{n\rightarrow\infty}(F_{n-1}/F_{n})=(\sqrt{5}-1)/2$, with $F_{n}$ being the $n$th Fibonacci numbers. For a finite system, one would choose the system size $L=F_{n}$ and $\alpha=F_{n-1}/F_{n}$ to impose the periodic boundary condition (PBC) for numerical diagonalization
of the tight-binding models in Eq. (\ref{eq:model1}) and Eq. (\ref{eq:model2}). 

\begin{figure*}[t]
	\centering
	\includegraphics[width=60mm,height=40mm]{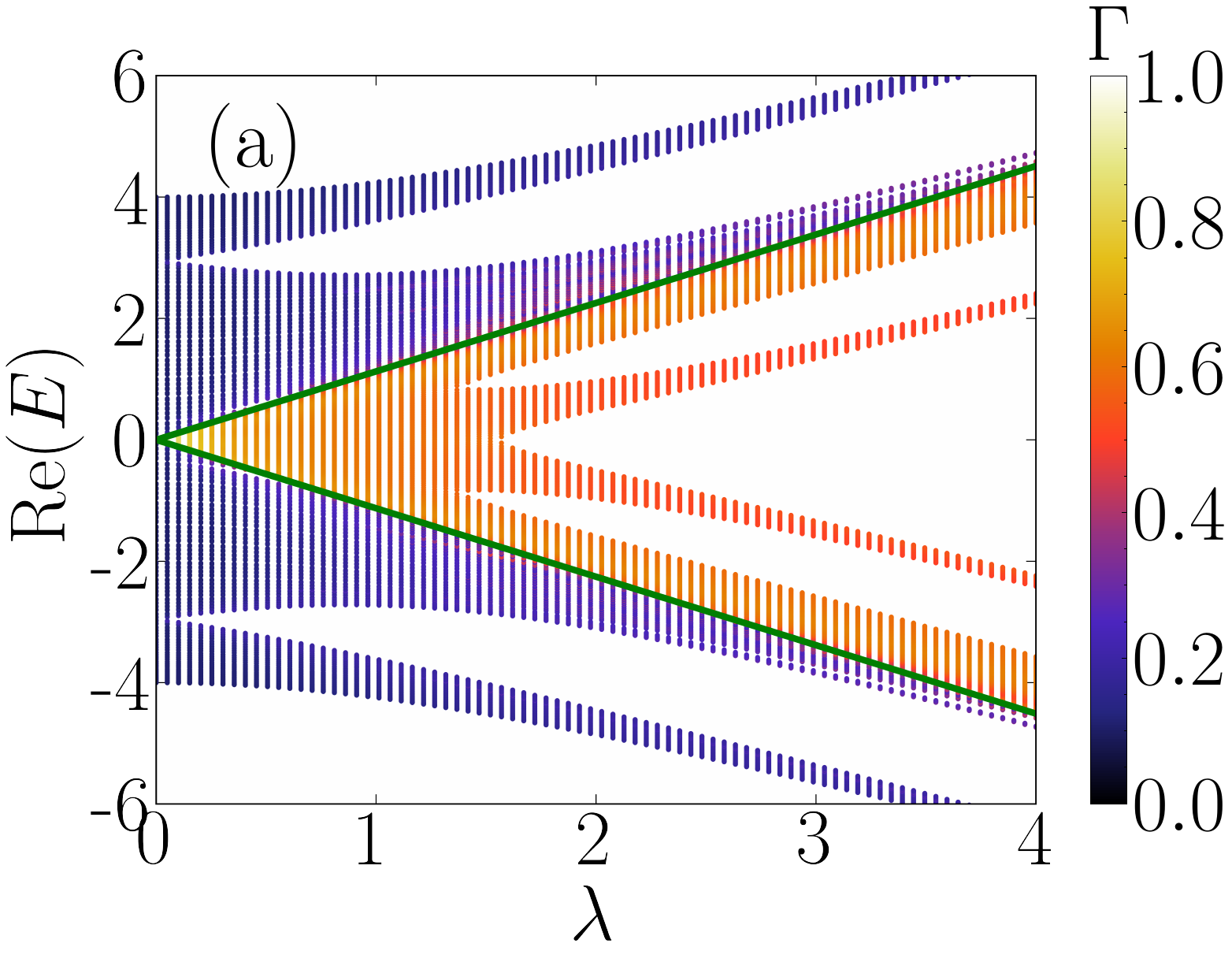}
	\includegraphics[width=57mm,height=38mm]{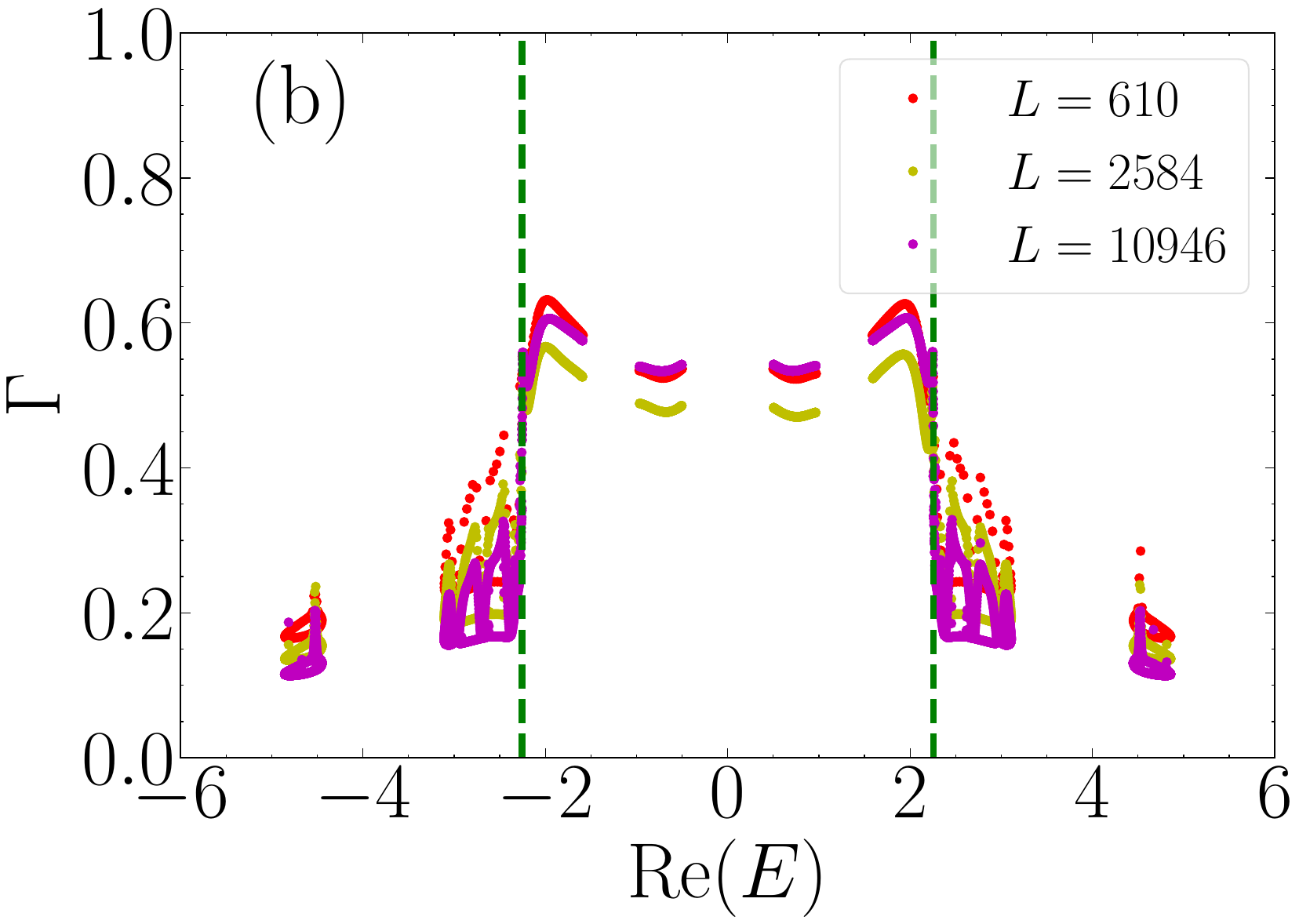}
	\includegraphics[width=57mm,height=38mm]{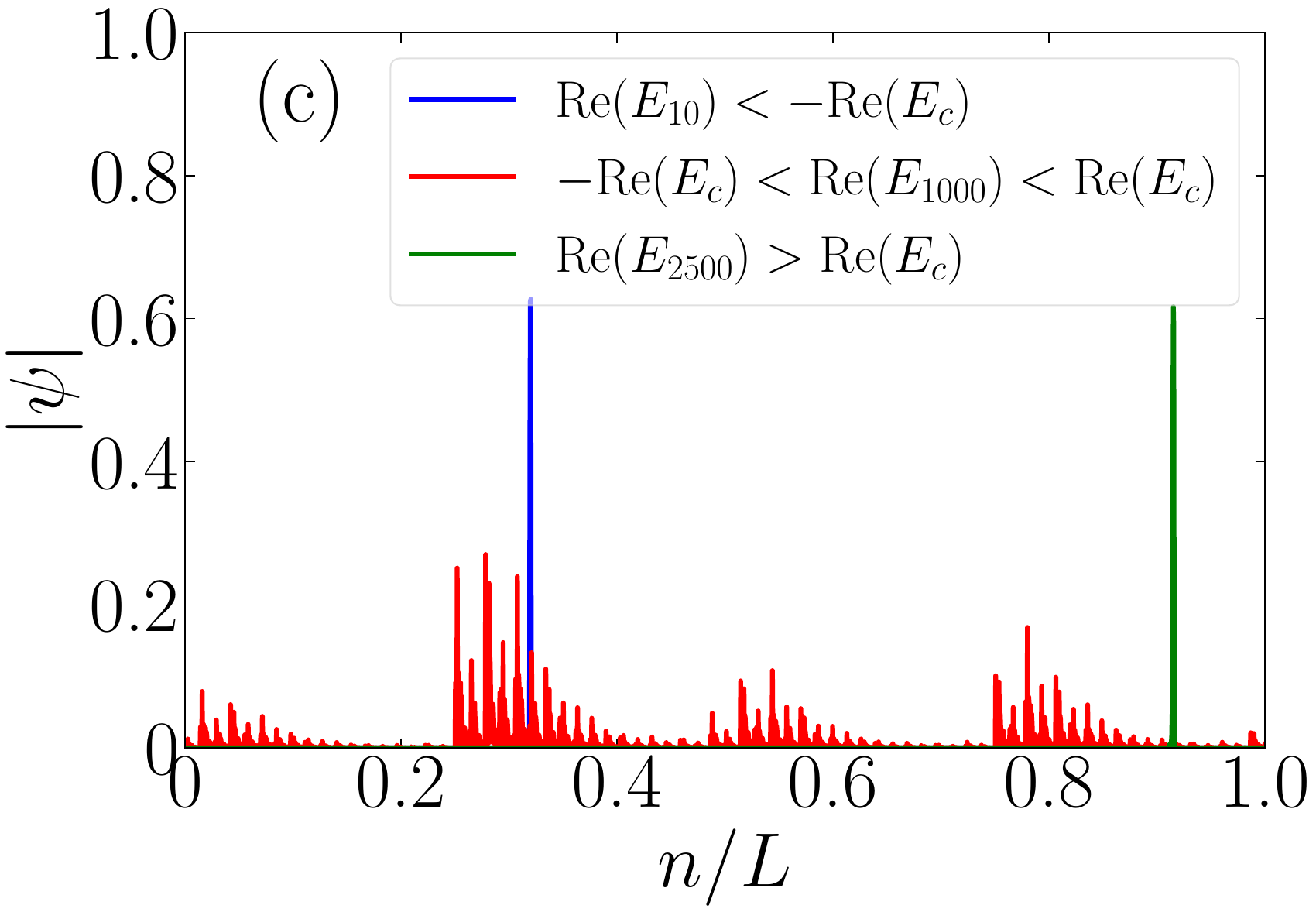}
	\caption{(a) Fractal dimension $ \Gamma $ of different eigenstates and the corresponding ${\rm{Re}}(E)$ as a function of $ \lambda $ for $L=2584$. The green solid lines represent the AMEs $|{\rm{Re}}(E_{c})|=\lambda \cosh(g)$. (b) $ \Gamma $ versus ${\rm{Re}}(E)$ with fixed $ \lambda=2.0 $ for $L=610$ (red dots), $L=2584$ (yellow dots), and $L=10946$ (magenta dots). The dashed lines denote the AMEs. (c) Spatial distributions of different typical eigenstates. $ |\psi| $ is the amplitude corresponding to the real spectrum ${\rm{Re}}(E_{10})<-{\rm{Re}}(E_{c})$, $-{\rm{Re}}(E_{c})<{\rm{Re}}(E_{1000})<{\rm{Re}}(E_{c})$, and ${\rm{Re}}(E_{2500})>{\rm{Re}}(E_{c})$ for $L=2584$ and $\lambda=2.0$, respectively. Critical states (red lines) and localized states (blue and green lines) are present. The other parameters are $V=1.0$ and $g=0.5$.}
	\label{fig2}
\end{figure*}

\begin{figure}[b]
	\centering
	\includegraphics[width=42mm,height=28mm]{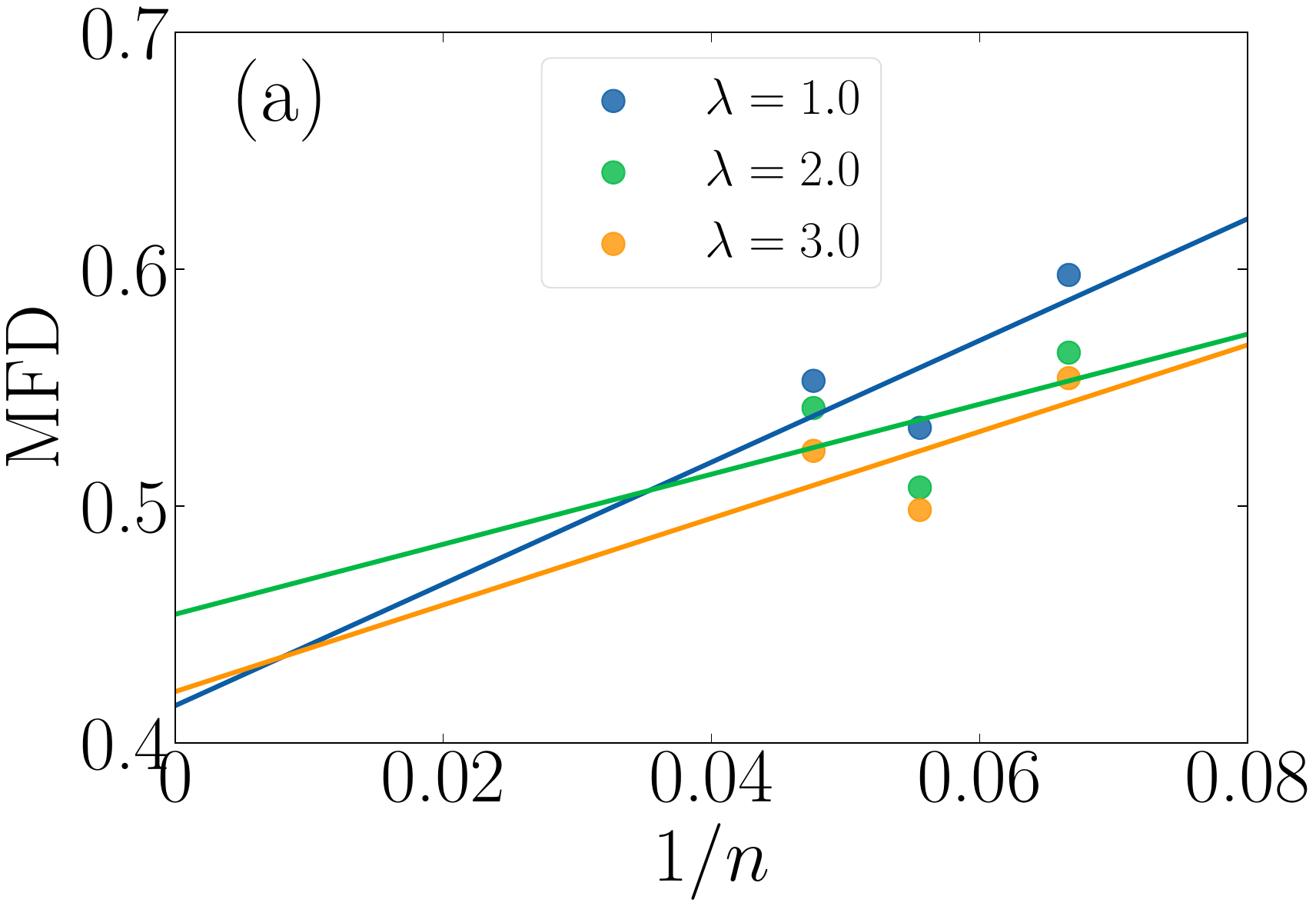}
	\includegraphics[width=42mm,height=28mm]{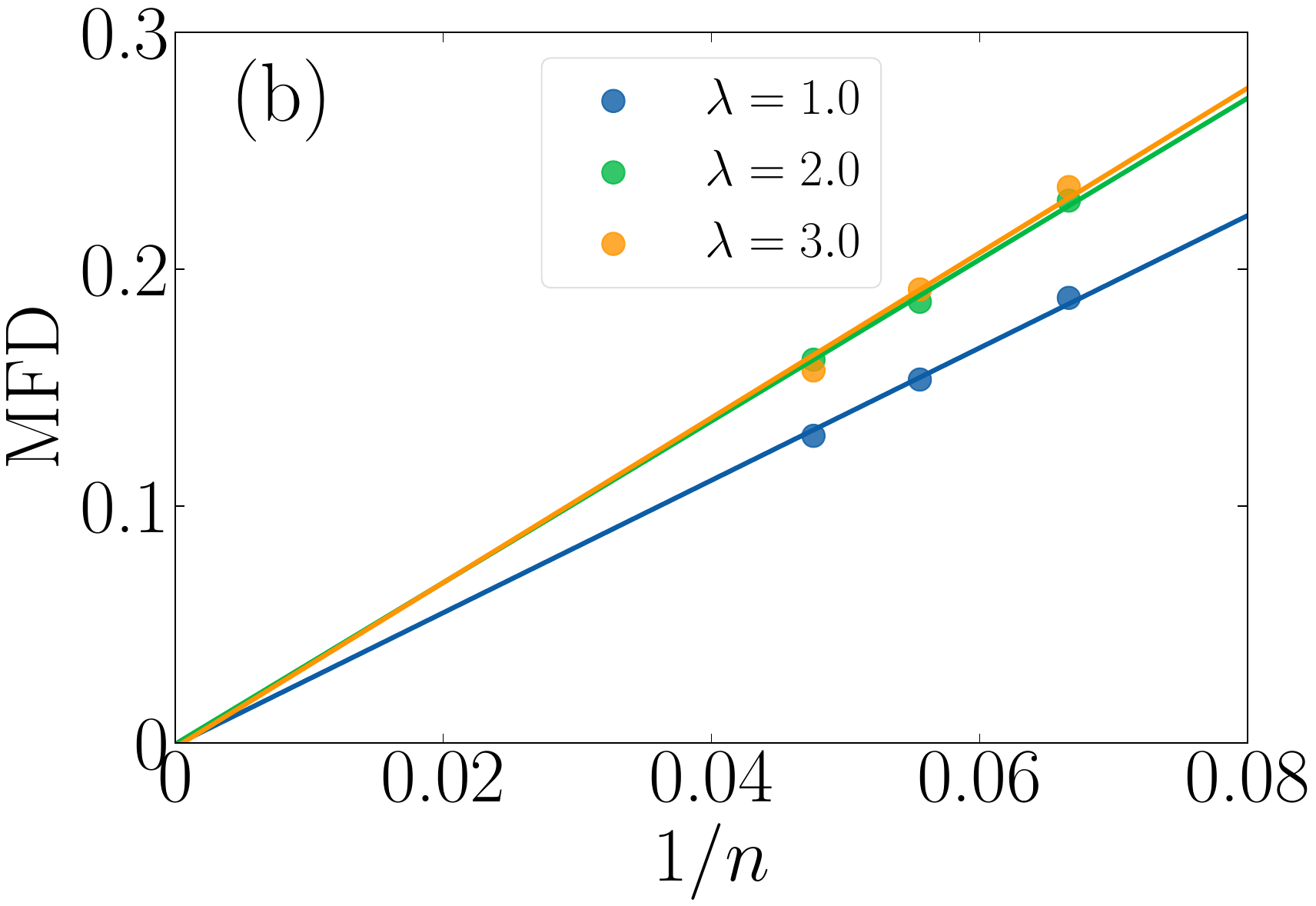}
	\includegraphics[width=42mm,height=28mm]{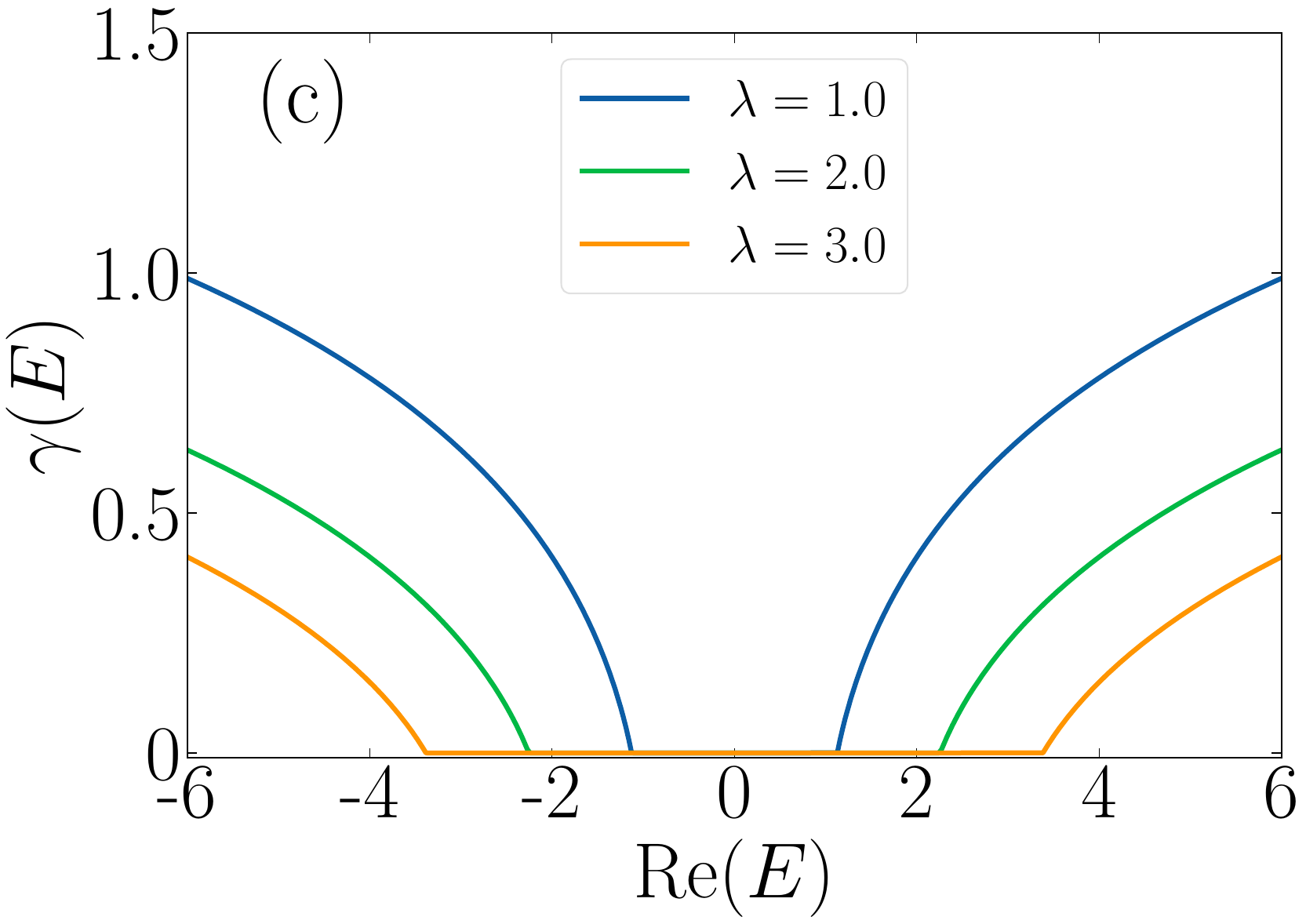}
	\includegraphics[width=42mm,height=28mm]{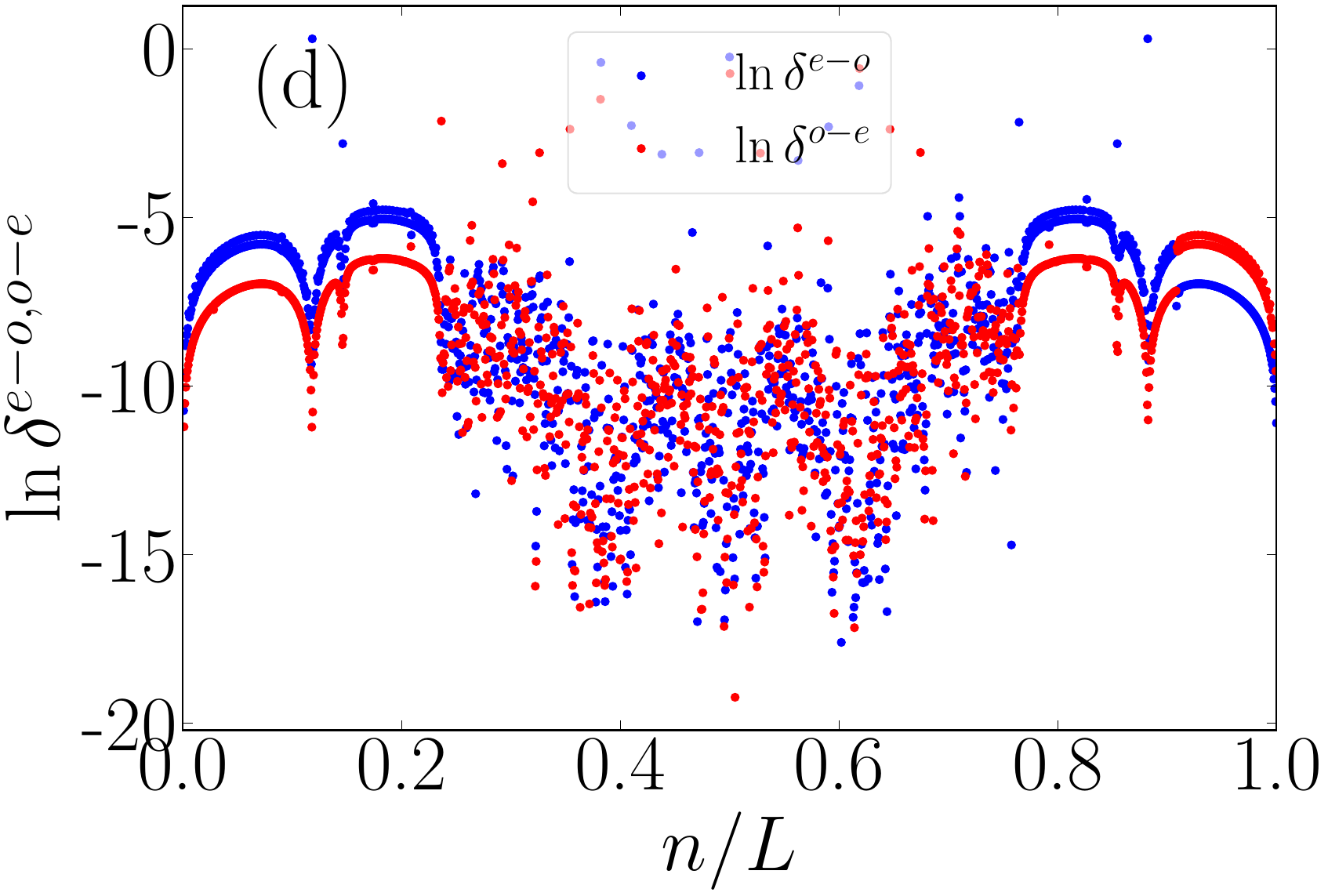}
	\caption{(a) and (b) show the MFD as a function $1/n$ for the critical and localized regions with different $\lambda$, respectively. Here $n$ is the index of the $n$th Fibonacci number $F_{n}$, and the system size is $L=F_{n}$. (C) The analytic LEs of the real part of eigenvalues for different $\lambda$.(d) The even-odd $\delta_{e-o}$ (blue dots) and odd-even $\delta_{o-e}$ (red dots) level spacings for the parameters $\lambda=2.0 $, $g=0.5 $, and the system size $L = 2584$.}
	\label{fig3}
\end{figure}

\section{Exact anomalous mobility edges in a quasiperiodic mosaic model}\label{sec:3}
In this section, we study model I which is featured by the mosaic AAH potentials of both hopping terms and on-site potentials. To comprehend the localization transition and the AMEs, we perform a similarity transformation on the Hamiltonian (\ref{eq:model1}) into the Hermitian Hamiltonian via a transformation $H_{\rm{I}}^{\prime}=S^{-1}_{\rm{I}}H_{\rm{I}}S_{\rm{I}}$, where the matrix $S_{\rm{I}}={\rm{diag}}\{1,1,r,r,...,r^{L/2},r^{L/2}\}$ and $r=e^{-g}$. Let $\psi^{\prime}$ denote the eigenstate of the transformed Hamiltonian $H_{\rm{I}}^{\prime}$, and $\psi$ be the eigenstate of the original Hamiltonian $H_{\rm{I}}$, it satisfies $\psi = S^{-1}_{\rm{I}}\psi^{\prime}$. Consequently, under the similarity transformation, for an extended eigenstate of  $H_{\rm{I}}^{\prime}$, $S^{-1}_{\rm{I}}$ localizes the wave function exponentially on the boundary, giving rise to the non-Hermitian skin effects~\cite{kunst2018biorthogonal,yao2018edge,gong2018topological}. Two localization lengths emerge on either side of the localized center for a localized state of the Hamiltonian $H_{\rm{I}}$. The AMEs and critical states of  $H_{\rm{I}}^{\prime}$ can be analytically derived by calculating the Lyapunov exponent (LE) using Avila's global theory\cite{avila2015global,wang2023exact}. Denote by $T_{n}(\phi)$ the transfer matrix of the Jacobi operator, and note that it can be expressed as:
\begin{equation}
T_2(\phi)=\frac{1}{\lambda M}
\begin{pmatrix}
E-M & -M \\
\lambda & 0
\end{pmatrix}
\begin{pmatrix}
E-M & -\lambda \\
M & 0
\end{pmatrix},
\end{equation}
where $M=2\cos(2\pi\alpha+\phi)$. Thus, the LE for an eigenstate with energy $E$ can be calculated via 
\begin{equation}
    \gamma_{\epsilon}(E)=\lim_{n\to\infty}\frac{1}{2\pi n}\int\ln\|T_{n}(\phi+i\epsilon)\|d\phi,
\end{equation}
where $\|\cdot\|$ represents the norm of the matrix and $\epsilon$ is imaginary part of complexified $\phi$, respectively. By a standard complexification procedure and using Avila's global theory, the LE is given by~\cite{zhou2023exact}
\begin{equation}
    \gamma_{0}(E)=\max\left\{\frac{1}{2}\ln|(|E|+\sqrt{E^{2}-\lambda^{2}})/\lambda|,0\right\}.
\end{equation}
For the Hamiltonian $H_{\rm{I}}$, we ultimately derive the LEs $\gamma(E)=\max\{\frac{1}{2}\ln|(|E|+\sqrt{E^{2}-\lambda^{2}})/\lambda|\pm g,0\}$. Let $\gamma(E)=0$, and we would have exact energy-dependent AMEs separating localized states and critical states, as indicated by
\begin{equation}\label{AME1}
|{\rm{Re}}(E_{c})|=\lambda \cosh(g).
\end{equation}
If $|{\rm{Re}}(E)|>\lambda \cosh(g)$, then $\gamma(E)>0$, the eigenenergy belongs to the point spectrum and the corresponding eigenstate is localized. Conversely, if $|{\rm{Re}}(E)|<\lambda \cosh(g)$, then $\gamma(E)=0$, the eigenstates can be extended or critical states, with the corresponding eigenenergy belonging to the absolutely continuous spectrum or singular continuous spectrum~\cite{avila2017spectral}, respectively. 

It is widely accepted that there are two primary methods for eliminating the presence of absolutely continuous spectrum (extended states): one involves introducing unbounded spectrum~\cite{liu2022anomalous,zhang2022lyapunov}, and the other involves introducing zeros in the hopping terms~\cite{simon1989trace,jitomirskaya2012analytic}. In our model I, there exists a sequence of sites $\{2n\}$ such that $t_{2n}\rightarrow0$ in the thermodynamic limit, thereby leading to the exclusion of extended states and the eigenstates associated with $|{\rm{Re}}(E)|\leq\lambda \cosh(g)$ being all critical states. In summary, the vanishing LEs and the absence of hopping coefficient zeros unambiguously determine the critical region for $|{\rm{Re}}(E)|\leq\lambda \cosh(g)$, while positive LEs delineate the localized region for $|{\rm{Re}}(E)|>\lambda \cosh(g)$. Therefore, Eq.(\ref{AME1}) signifies the critical energies separating localized and critical states, manifesting the AMEs.

To numerically verify the analytical results we obtained, we can use fractal dimension (FD) and energy spectrum statistics to identify the extended, localized, and critical states~\cite{evers2008anderson,qi2023multiple}. For an arbitrary given $m$-th eigenstate $|\Psi_{m}\rangle=\sum_{n=1}^{L}\psi_{m,n}a_{n}^{\dagger}|0\rangle$, the inverse participation ratio (IPR) being $\mathrm{IPR}=\sum_{j}|\psi_{m,j}|^{4}$. Consequently, the FD $\Gamma=-\lim_{L\rightarrow\infty}\ln(\mathrm{IPR})/\ln(L)$. In the thermodynamics limit, the $\Gamma$ approaches $1$ for extended states and $0$ for localized states, whereas $0<\mathrm{FD}<1$ for critical states. Figure~\ref{fig2}(a) illustrates the $\Gamma$ as a function of $\lambda$ for for various eigenvalues ${\rm{Re}}(E)$. The green solid lines, originating from the band center, represent the AMEs $|{\rm{Re}}(E_{c})|=\lambda \cosh(g)$, across which $\Gamma$ varies from approximately $0.5$ to $0.1$, highlighting a critical-to-localization transition predicted by the analytic results. We further present the spatial distributions of three typical eigenstates in Fig.~\ref{fig2}(c), where the eigenstates corresponding to real eigenvalues ${\rm{Re}}(E_{10})<-{\rm{Re}}(E_{c})$ or ${\rm{Re}}(E_{2500})>{\rm{Re}}(E_{c})$ are localized, whereas the eigenstate with real eigenvalue $-{\rm{Re}}(E_{c})<{\rm{Re}}(E_{1000})<{\rm{Re}}(E_{c})$ is critical.  Notably, in Fig.~\ref{fig2}(b), we fix the parameters $\lambda=2.0$, $g=0.5$ and depict $\Gamma$ as a function of the corresponding eigenvalues ${\rm{Re}}(E)$ for various system sizes $L$. The green dashed lines in the figure represent the AMEs ${\rm{Re}}(E_{c}) \simeq \pm 2.26$. One can observe that in Fig.~\ref{fig2}(b), the $\Gamma$ tends to $0$ for all eigenstates in energy zones with $|{\rm{Re}}(E)|> 2.26$ as the system size increases, suggesting that these eigenstates are localized. In contrast, in energy zones with $|{\rm{Re}}(E)|\leq 2.26$, is nearly independent of the system size and differs significantly from $ 0$ and $1$, approaching $0.5$ magnitude, indicating that these eigenstates are critical. A more meticulous finite-size scaling for mean fractal dimension (MFD) can be found in Figs.~\ref{fig3} (a) and (b), where it is shown that the MFD of the critical zone converges to a finite value, whereas the MFD of the localized zone tends to $0$ as the system size grows.  In Figure~\ref{fig3} (c), we also plot the LEs of the $H_{\rm{I}}$ for different parameters $\lambda$.

To more clearly distinguish between extended, critical, and localized states, we define the even-odd (odd-even) level spacings of the eigenvalues~\cite{deng2019one} as $\delta_{n}^{e-o}={\rm{Re}}(E_{2n})-{\rm{Re}}(E_{2n-1})\left(\delta_{n}^{o-e}={\rm{Re}}(E_{2n+1})-{\rm{Re}}(E_{2n})\right)$. ${\rm{Re}}(E_{2n})$ and ${\rm{Re}}(E_{2n-1})$ denote the even and odd eigenenergies in ascending order of the real eigenenergy spectrum, respectively. In the extended region, the eigenenergy spectrum for the system is nearly doubly degenerate, leading to the vanishing of  $\delta_{n}^{e-o}$. Consequently, a significant gap exists between $\delta_{n}^{e-o}$ and $\delta_{n}^{o-e}$. In the localized region, $\delta_{n}^{e-o}$ and $\delta_{n}^{o-e}$ are almost the same and the gap disappears. In the critical region, $\delta_{n}^{e-o}$ and $\delta_{n}^{o-e}$ exhibit scattered distribution behavior, which is distinct from extended and localized phases. As depicted in Fig~\ref{fig3} (d), our numerical results reveal that the central eigenvalues correspond to critical states, while the energy spectra at the two boundaries are localized states.

\begin{figure*}[t]
	\centering
	\includegraphics[width=60mm,height=40mm]{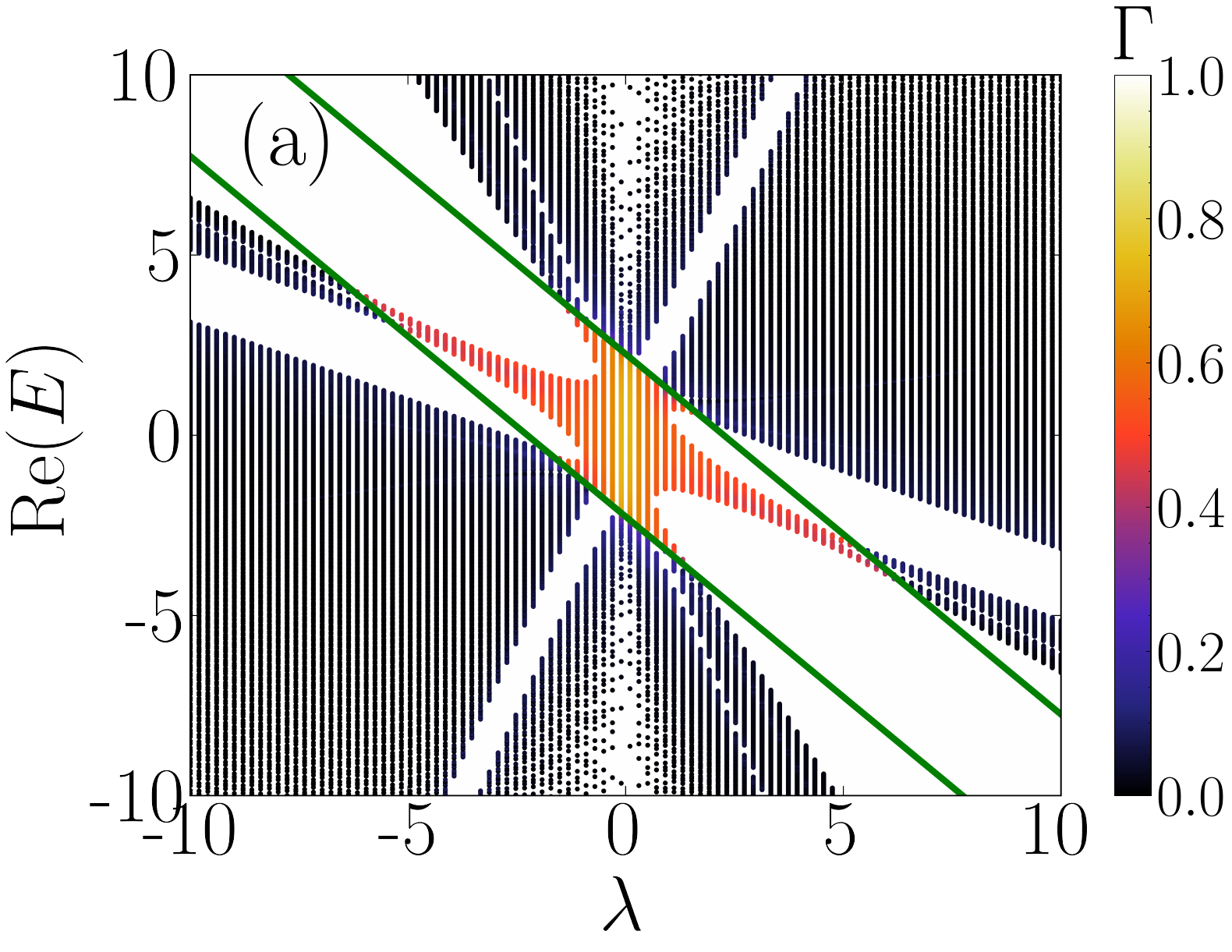}
	\includegraphics[width=57mm,height=38mm]{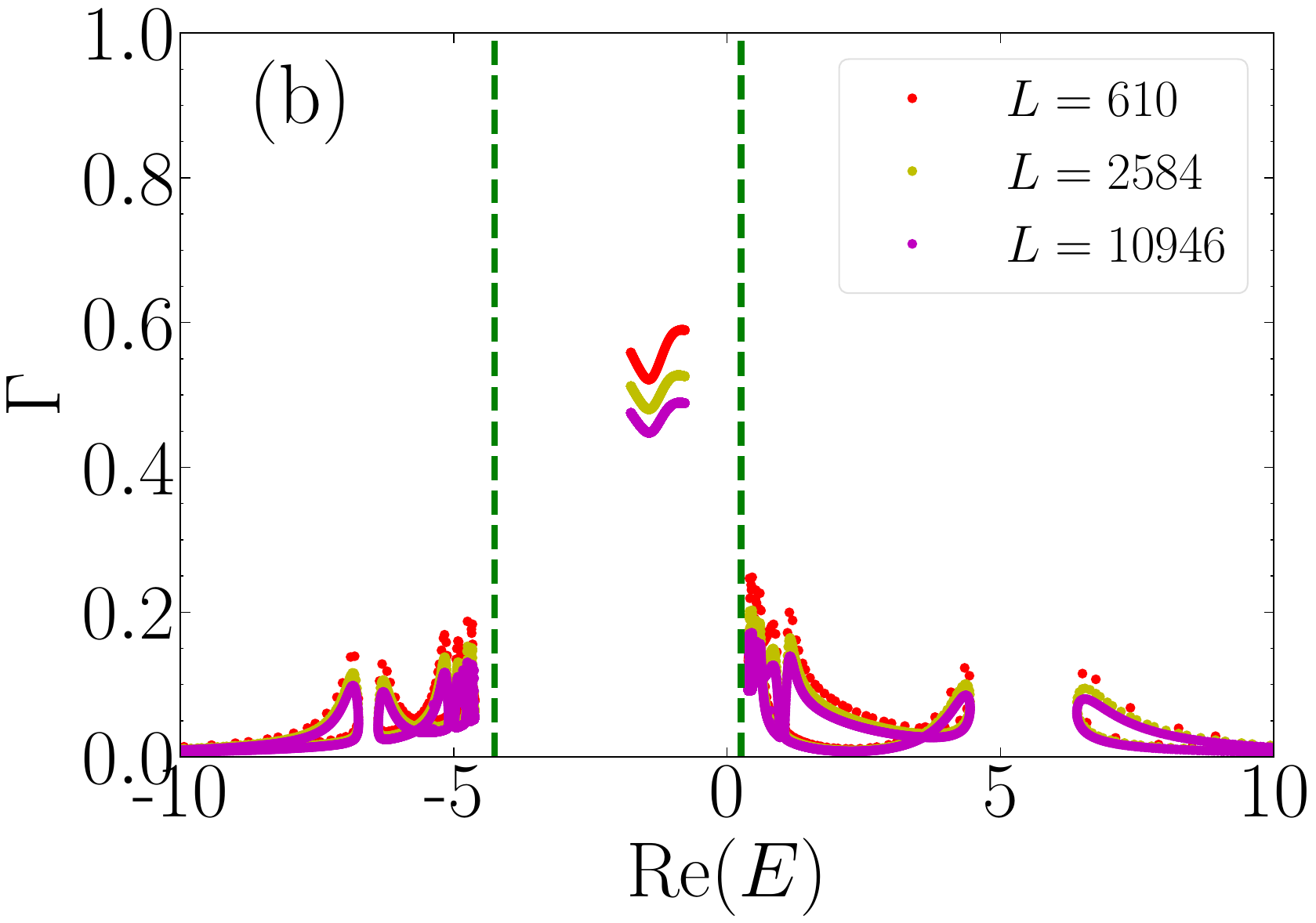}
	\includegraphics[width=57mm,height=38mm]{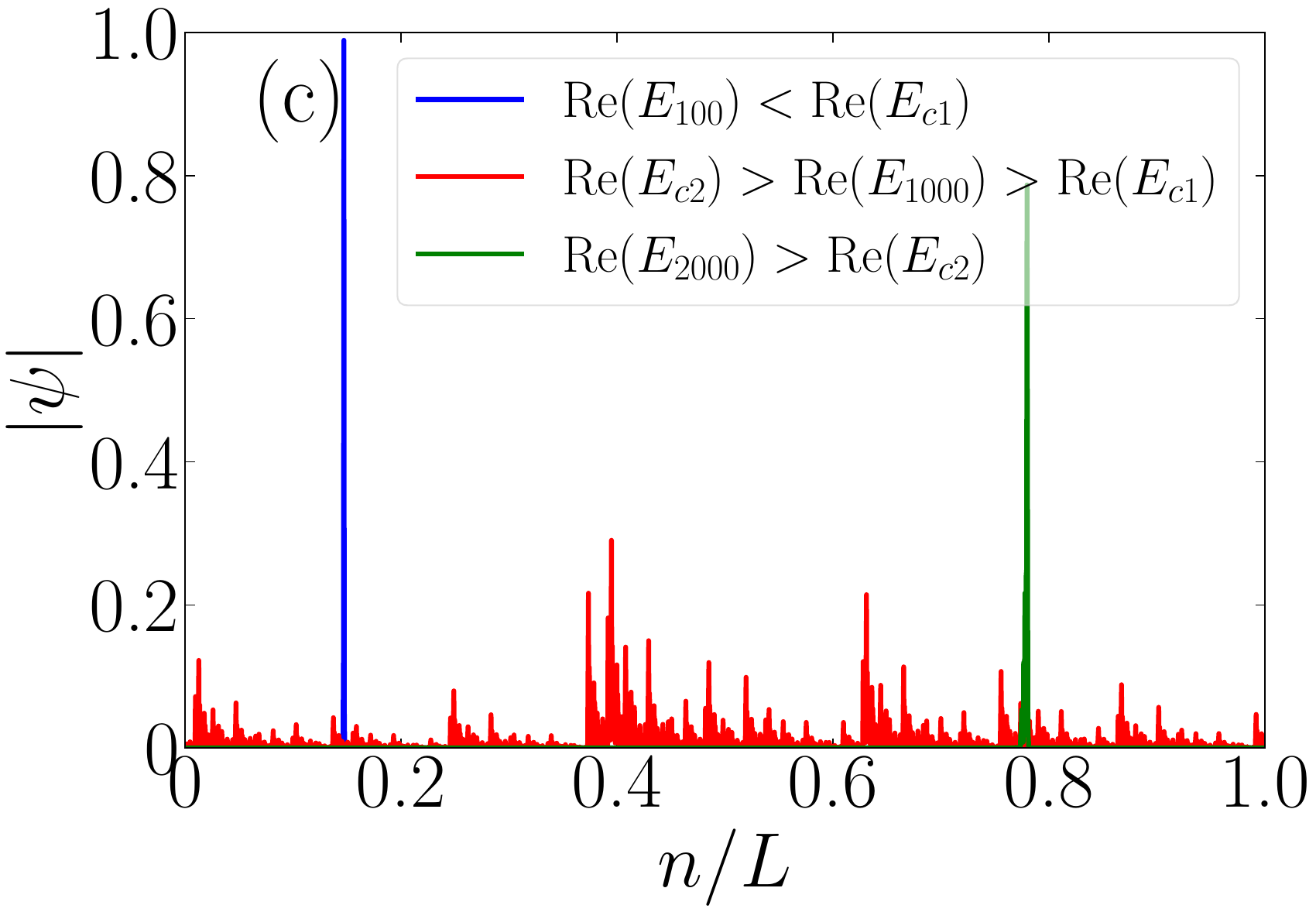}
	\caption{(a) Fractal dimension $ \Gamma $ of different eigenstates and the corresponding ${\rm{Re}}(E)$ as a function of $ \lambda $ for $L=2584$ and $g=0.5$. The green solid lines represent the AMEs ${\rm{Re}}(E_{c})=(\pm 2b\cosh(g)-2\lambda)/b$. (b) $ \Gamma $ versus ${\rm{Re}}(E)$ with fixed $ \lambda=2.0 $ for $L=610$ (red dots), $L=2584$ (yellow dots), and $L=10946$ (magenta dots). The dashed lines denote the AMEs. (c) Spatial distributions of different typical eigenstates. $ |\psi| $ is the amplitude corresponding to the real spectrum ${\rm{Re}}(E_{100})<-{\rm{Re}}(E_{c})$, $-{\rm{Re}}(E_{c})<{\rm{Re}}(E_{1000})<{\rm{Re}}(E_{c})$, and ${\rm{Re}}(E_{2000})>{\rm{Re}}(E_{c})$ for $L=2584$, respectively. Critical states (red line) and localized states (blue and green lines) are present. The other parameters are $t=1.0$, $b=2.0$, and $g=0.5$.}
	\label{fig4}
\end{figure*}

\begin{figure}[b]
	\centering
	\includegraphics[width=42mm,height=28mm]{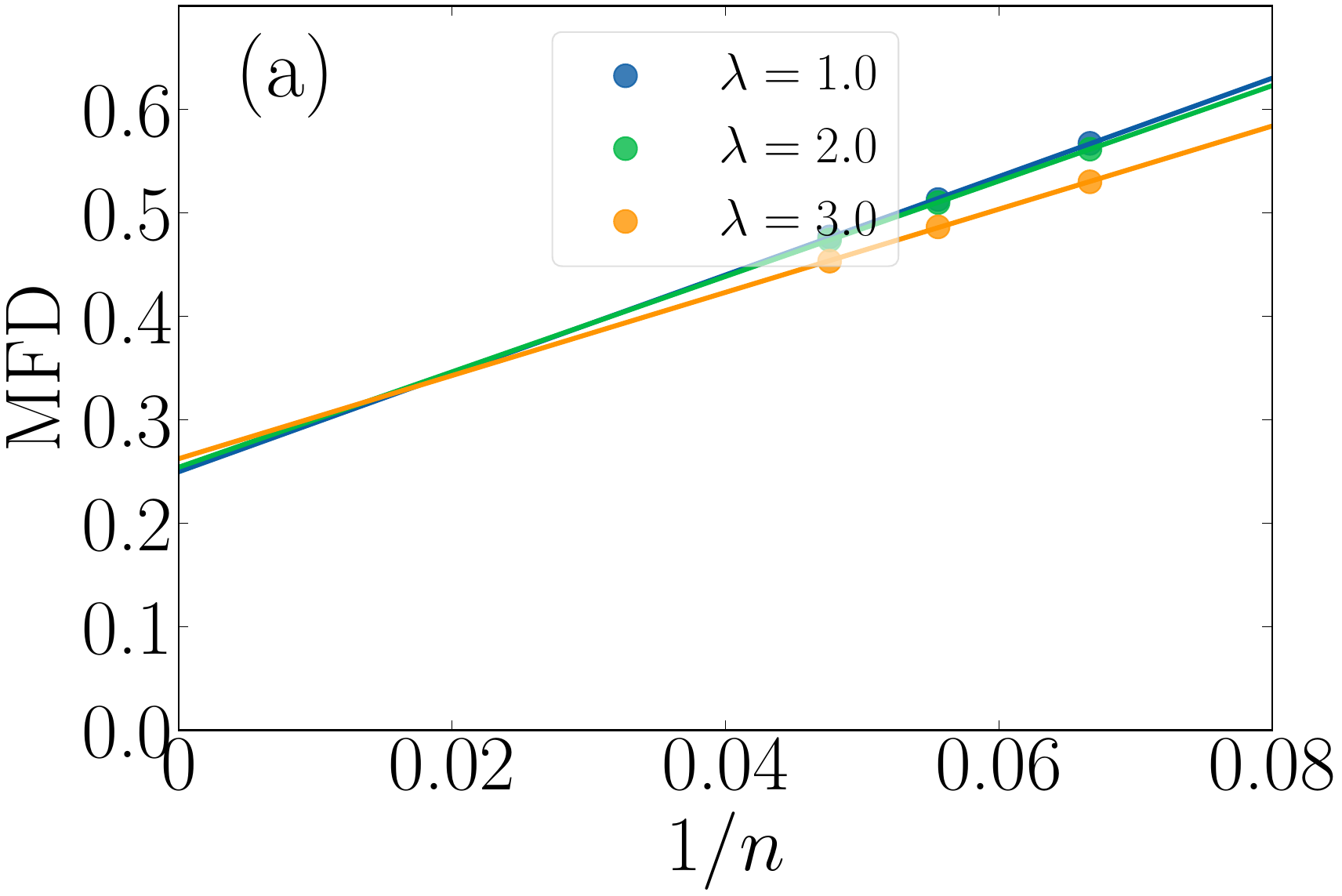}
	\includegraphics[width=42mm,height=28mm]{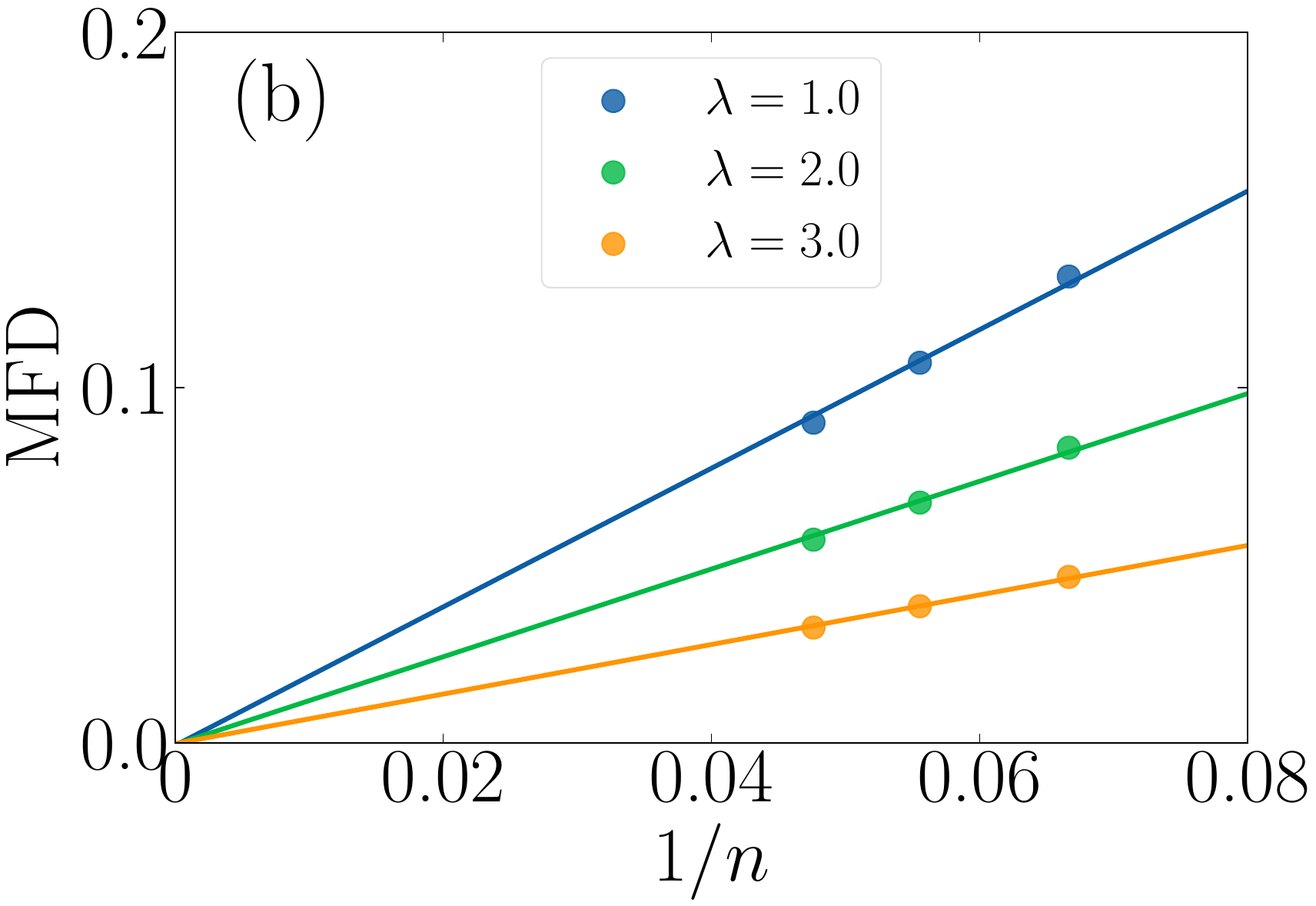}
	\includegraphics[width=42mm,height=28mm]{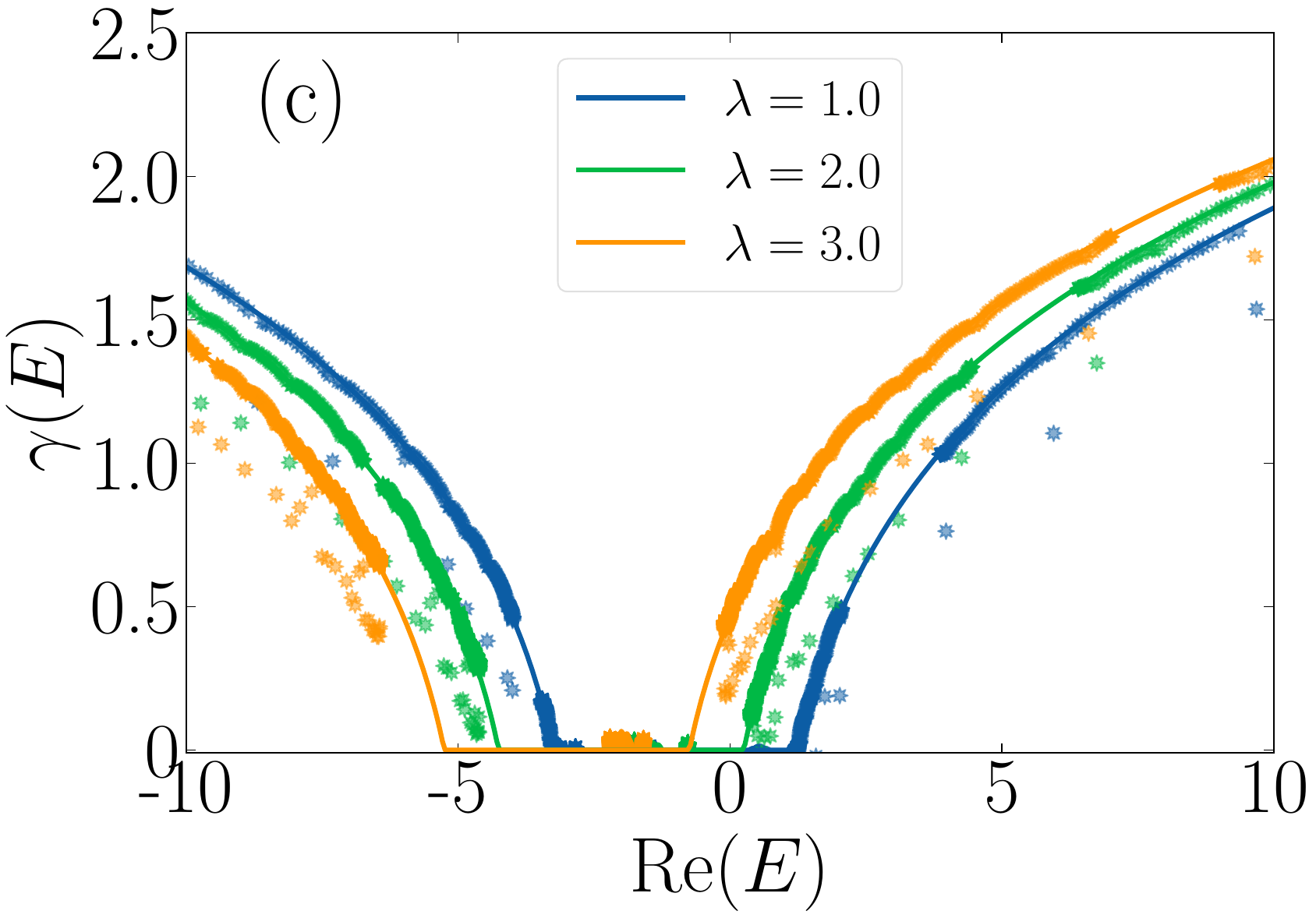}
	\includegraphics[width=42mm,height=28mm]{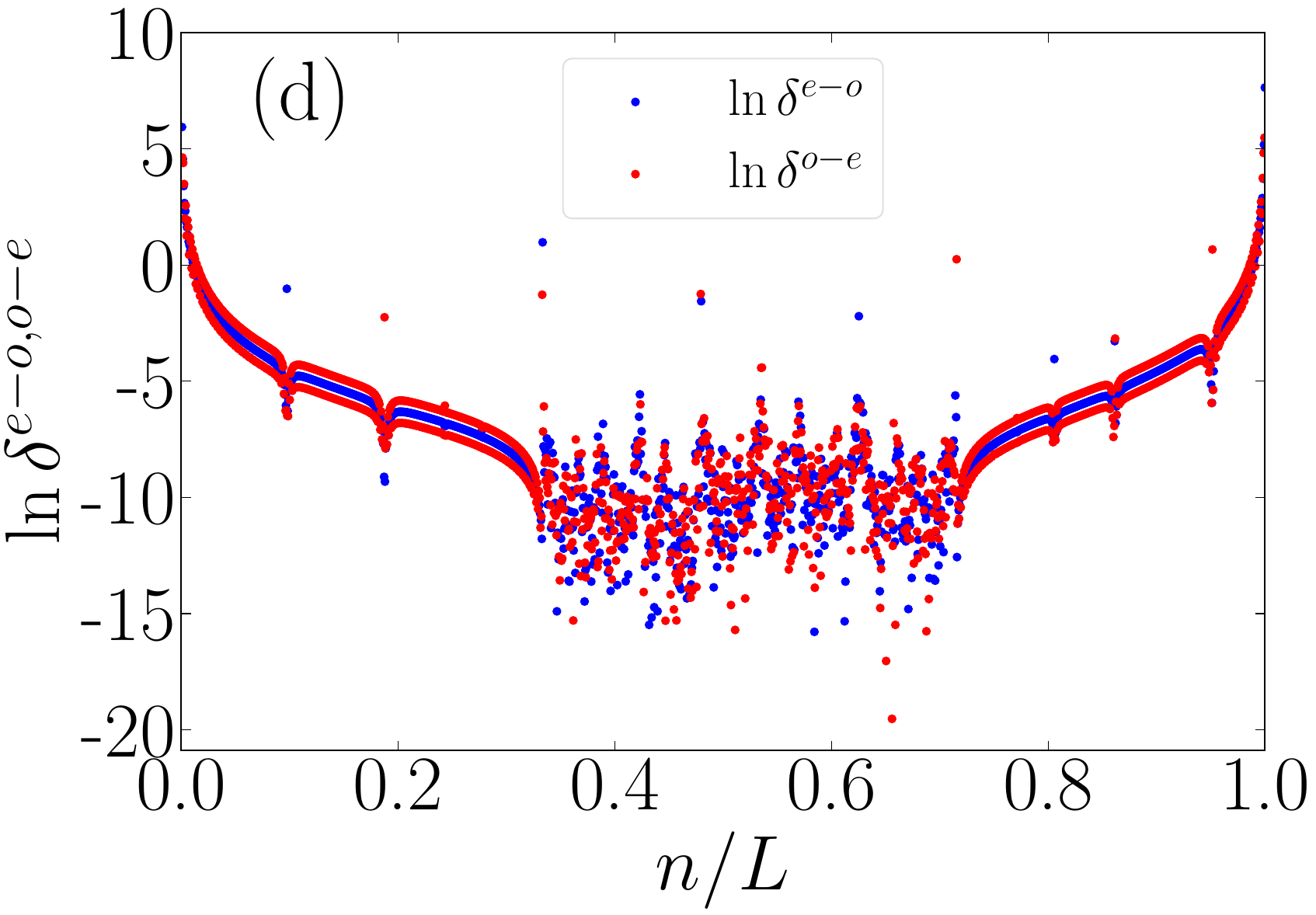}
	\caption{(a) and (b) show the MFD as a function $1/n$ for the critical and localized regions with different $\lambda$, respectively. Here $n$ is the index of the $n$th Fibonacci number $F_{n}$, and the system size is $L=F_{n}$. (C) The numerical LEs (dots) and analytic LEs (lines) of the real part of eigenvalues for different $\lambda$. (d) The even-odd $\delta_{e-o}$ (blue dots) and odd-even $\delta_{o-e}$ (red dots) level spacings for the system size $L = 2584$. The other parameters are $b=2$ and $g=0.5$.}
	\label{fig5}
\end{figure}

\section{Exact anomalous mobility edges in an unbounded quasiperiodic model}\label{sec:4}
In this section, we investigate model II Hamiltonian (\ref{eq:model2}), which exhibits nonreciprocal hopping and GPD potential (\ref{eq5}) with $\abs{b}\geqslant 1$. The LE can characterize the localized properties of eigenstates. We present the transfer matrix method~\cite{davids1995lyapunov} and its relation to the LE. Initially, we transform given the Hamiltonian (\ref{eq:model2}) into the Hermitian Hamiltonian using a similar transformation $H_{\rm{II}}^{\prime}=S^{-1}_{\rm{II}}H_{\rm{II}}S_{\rm{II}}$. Then, starting from the eigenstate of the transformed Hamiltonian, we derive the LE of the original Hamiltonian. The similar matrix $S_{\rm{II}}={\rm{diag}}\{1,r,r^2,,...,r^{L}\}$ is defined with  $r=e^{-g}$. Let $\psi^{\prime}$ denote the eigenstate of the transformed Hamiltonian $H_{\rm{II}}^{\prime}$, and since $\psi$ is the eigenstate of the original Hamiltonian $H_{\rm{II}}$,  it follows that $\psi = S^{-1}_{\rm{II}}\psi^{\prime}$. Assuming the system to be a half-infinite lattice with left-hand end sites $n=0$ and $n=1$, the LE of $H_{\rm{II}}^{\prime}$ can be determined using the transfer matrix method. For instance, by starting with $\psi^{\prime}(0)$ and $\psi^{\prime}(1)$ of the left-hand end sites, the wave function can be derived through the relation
\begin{align}\label{V}
\Psi^{\prime}(n)=T(n)T(n-1)...T(2)T(1)\Psi^{\prime}(0)
\end{align}
where matrix
\begin{align}\label{V}
T(n)\equiv
\left(\begin{array}{ccc}
E-\frac{ 2\lambda\cos(2\pi\alpha n+\phi)}{1-b\cos(2\pi\alpha n+\phi)} &-1  \\
1&0\\
\end{array}\right).
\end{align}
and
\begin{align}\label{V}
\Psi^{\prime}(n)\equiv\left(\begin{array}{ccc}
\psi^{\prime}(n+1)  \\
\psi^{\prime}(n)\\
  \end{array}\right).
\end{align}
Viewing the aforementioned equation as an evolutionary equation of a dynamical system, $\psi(0)$ and $\psi(1)$ act as the initial conditions. Given a real number $E$, as $n$ increases, one may assume that the wave function grows approximately according to an exponential law, i.e.,$\psi^{\prime}(n)\sim e^{\gamma^{\prime}(E) n}$,  as $n\rightarrow \infty$, where $\gamma^{\prime}(E)\geq0$ is the LE. If the parameter $E$ is not an eigenenergy of $H_{\rm{II}}^{\prime}$, the LE would be positive, $\gamma^{\prime}(E)>0$. Conversely, if the parameter $E$ is an eigenenergy of $H$, the LE can be zero or positive.
For extended or critical states, the LE $\gamma^{\prime}(E)\equiv0$. Conversely, for localized states, the LE $\gamma^{\prime}(E)>0$. Therefore, the LE of $H_{\rm{II}}^{\prime}$ can be expressed as~\cite{zhang2022lyapunov} 
\begin{align}\label{V}
&\gamma^{\prime}(E)=\lim_{L \rightarrow \infty }\frac{\ln(|\Psi^{\prime}(L)|/|\Psi^{\prime}(0)|)}{L} \notag\\
&=\lim_{L\rightarrow \infty}\frac{\ln(|T(L)T(L-1)...T(2)T(1)\Psi^{\prime}(0)|/|\Psi^{\prime}(0)|)}{L}
\end{align}
where $L$ is the system size and $|\Psi^{\prime}(n)|=\sqrt{|\psi^{\prime}(n+1)|^2+|\psi^{\prime}(n)|^2}$. In accordance with Refs.~\cite{wang2023exact,wang2021duality}, we complexify the phase $\phi \rightarrow \phi+i\epsilon$ and take advantage of the ergodicity of the map $\phi \rightarrow 2\pi\alpha n+\phi$. Consequently, we can express the LE as an integral over the phase $\phi$ as follows:
\begin{equation}
    \gamma^{\prime}_{\epsilon}(E)=\lim_{n\to\infty}\frac{1}{2\pi n}\int\ln\|T_{n}(\phi+i\epsilon)\|d\phi,
\end{equation}
where $\|\cdot\|$ signifies the norm of the matrix, and $\epsilon$ represents the imaginary component of the complexified $\phi$. Utilizing a standard complexification procedure and incorporating Avila's global theory, the LE is derived as
\begin{equation}
    \gamma^{\prime}_{0}(E)=\max\left\{\ln \left| \frac{|bE+2\lambda|+\sqrt{(bE+2\lambda)^{2}-4b^{2}}}{2b} \right |,0\right\}.
\end{equation}
As a result, for the Hamiltonian $H_{\rm{II}}$ and thanks to the similarity transformation $\psi = S^{-1}_{\rm{II}}\psi^{\prime}$, we ultimately determine the LEs $\gamma(E)=\max\{\gamma^{\prime}_{0}(E)\pm g,0\}$. Upon setting $\gamma(E)=0$, we would have exact energy-dependent AMEs that separate localized states and critical states, yielding 
\begin{equation}\label{AME2}
{\rm{Re}}(E_{c})=[\pm 2b\cosh(g)-2\lambda]/b.
\end{equation}
If ${\rm{Re}}(E)>[2b\cosh(g)-2\lambda]/b$ or ${\rm{Re}}(E)<[-2b\cosh(g)-2\lambda]/b$, then $\gamma(E)>0$, the eigenenergy belongs to the point spectrum and the corresponding eigenstate is localized. If $[-2b\cosh(g)-2\lambda]/b<{\rm{Re}}(E)<[2b\cosh(g)-2\lambda]/b$, then $\gamma(E)=0$, the eigenstates can either be extended or critical states and the corresponding eigenenergy belong to absolutely continuous spectrum or singular continuous spectrum, respectively. It is known that, for our model II, the Hamiltonian $H_{\rm{II}}$  has an unbounded spectrum, and the eigenstates associated with $\gamma(E)=0$ are all critical states. Thus, Eq.(\ref{AME2}) marks critical energies separating localized states and critical states, manifesting AMEs.

To validate the analytical outcomes we have derived, we perform exact diagonalization of $H_{\rm{II}}$ under PBC and employ FD and energy spectrum statistics to distinguish between critical and localized states. As illustrated in Fig.~\ref{fig5}(a), we display the FD $\Gamma$ as a function of $\lambda$ for various eigenvalues ${\rm{Re}}(E)$ at the parameter $g=0.5$. The green solid lines represent the AMEs Eq.(\ref{AME2}). The $\Gamma$ magnitude between the two lines is approximately $0.5$, signifying critical zones, whereas the  $\Gamma$ magnitude outside the two lines is close to $0$, denoting localized zones. Subsequently, in Fig.~\ref{fig5}(b), we fix the parameters $\lambda=2.0$ and $g=0.5$ and present the $\Gamma$ as a function of the corresponding eigenvalues ${\rm{Re}}(E)$ for different systems sizes $L$. The green dashed lines in the figure represent the AMEs ${\rm{Re}}(E_{c1}) \simeq -4.26$ and ${\rm{Re}}(E_{c2}) \simeq 0.26$. One can observe that in Fig.~\ref{fig5}(b), the $\Gamma$ tends to $0$ for all eigenstates in energy zones with ${\rm{Re}}(E)<{\rm{Re}}(E_{c1})$ or ${\rm{Re}}(E)>{\rm{Re}}(E_{c2})$ with the system size increasing, suggesting that these eigenstates are localized. In contrast, in energy zones with ${\rm{Re}}(E_{c2})>{\rm{Re}}(E)>{\rm{Re}}(E_{c1})$, the $\Gamma \simeq 0.5$ magnitude is far different from 0 and 1, and nearly independent of the system size, indicating that these eigenstates are critical. We further present the spatial distributions of several typical eigenstates in Fig.~\ref{fig5}(c), where the eigenstate of a real eigenvalue ${\rm{Re}}(E_{100})<{\rm{Re}}(E_{c1})$ or ${\rm{Re}}(E_{2000})>{\rm{Re}}(E_{c2})$ is localized, whereas the eigenstate of a real eigenvalue ${\rm{Re}}(E_{c1})<{\rm{Re}}(E_{1000})<{\rm{Re}}(E_{c2})$ is critical. Further, finite-size scaling analysis for MFD of various parameters $\lambda$ can be found in Figs.~\ref{fig5} (a) and (b). We observe that the MFD of the critical zone approaches a finite value of $0.25$, while that of the localized zone tends to be $0$ as the system size increases. In Figure~\ref{fig5} (c), we also plot the LEs of the $H_{\rm{II}}$ for different parameters $\lambda$, and the numerical results align with the analytical LE $\gamma(E)$. Finally, considering the even-odd (odd-even) level spacings of the eigenvalues as in the previous section, we define $\delta_{n}^{e-o}={\rm{Re}}(E_{2n})-{\rm{Re}}(E_{2n-1})\left(\delta_{n}^{o-e}={\rm{Re}}(E_{2n+1})-{\rm{Re}}(E_{2n})\right)$. ${\rm{Re}}(E_{2n})$ and ${\rm{Re}}(E_{2n-1})$ denote the even and odd eigenenergies in ascending order of the real eigenenergy spectrum, respectively. It is known that for localized states, $\delta_{n}^{e-o}$ and $\delta_{n}^{o-e}$ are almost the same, and the gap no longer exists. For the critical states, $\delta_{n}^{e-o}$ and $\delta_{n}^{o-e}$ have scatter-distributed behavior. As depicted in Fig.~\ref{fig5} (d), for the system size $L=2584$ and the parameters $\lambda=2$ and $g=0.5$, our numerical results indicate that the center eigenvalues are the critical states, while the energy spectra at the two boundaries are localized states.

\begin{figure}[t]
	\centering
	\includegraphics[width=42mm,height=28mm]{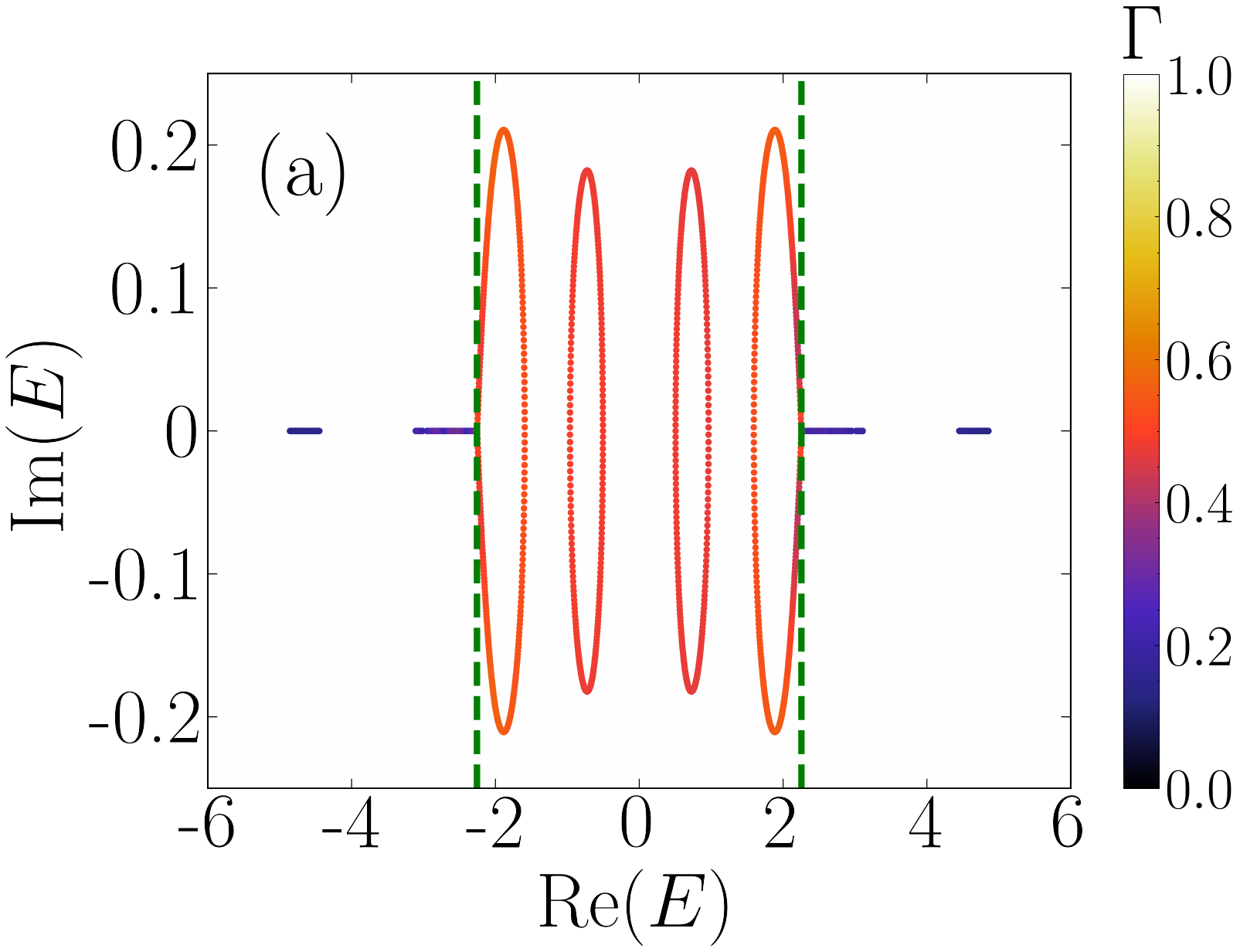}
	\includegraphics[width=42mm,height=28mm]{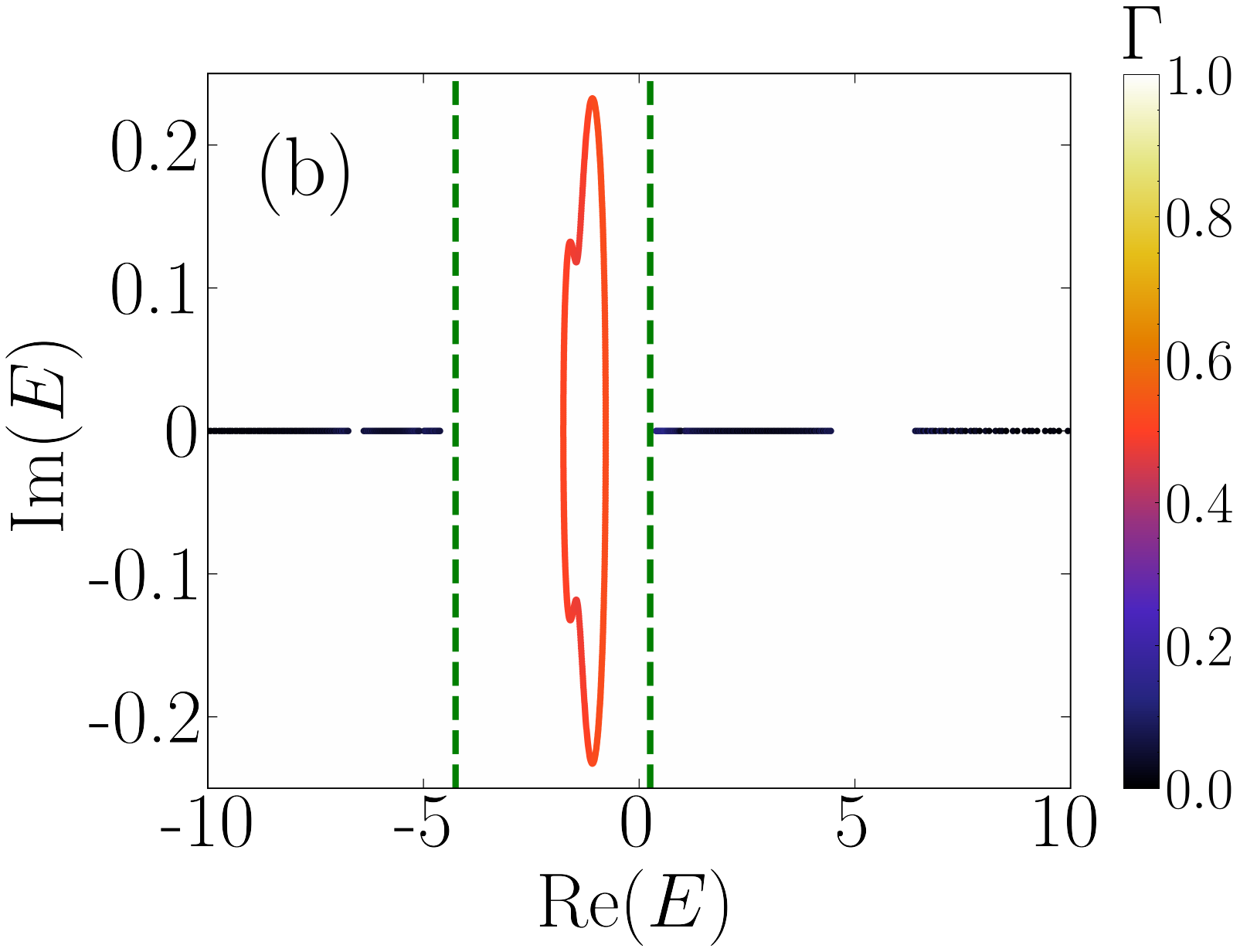}
	\includegraphics[width=42mm,height=28mm]{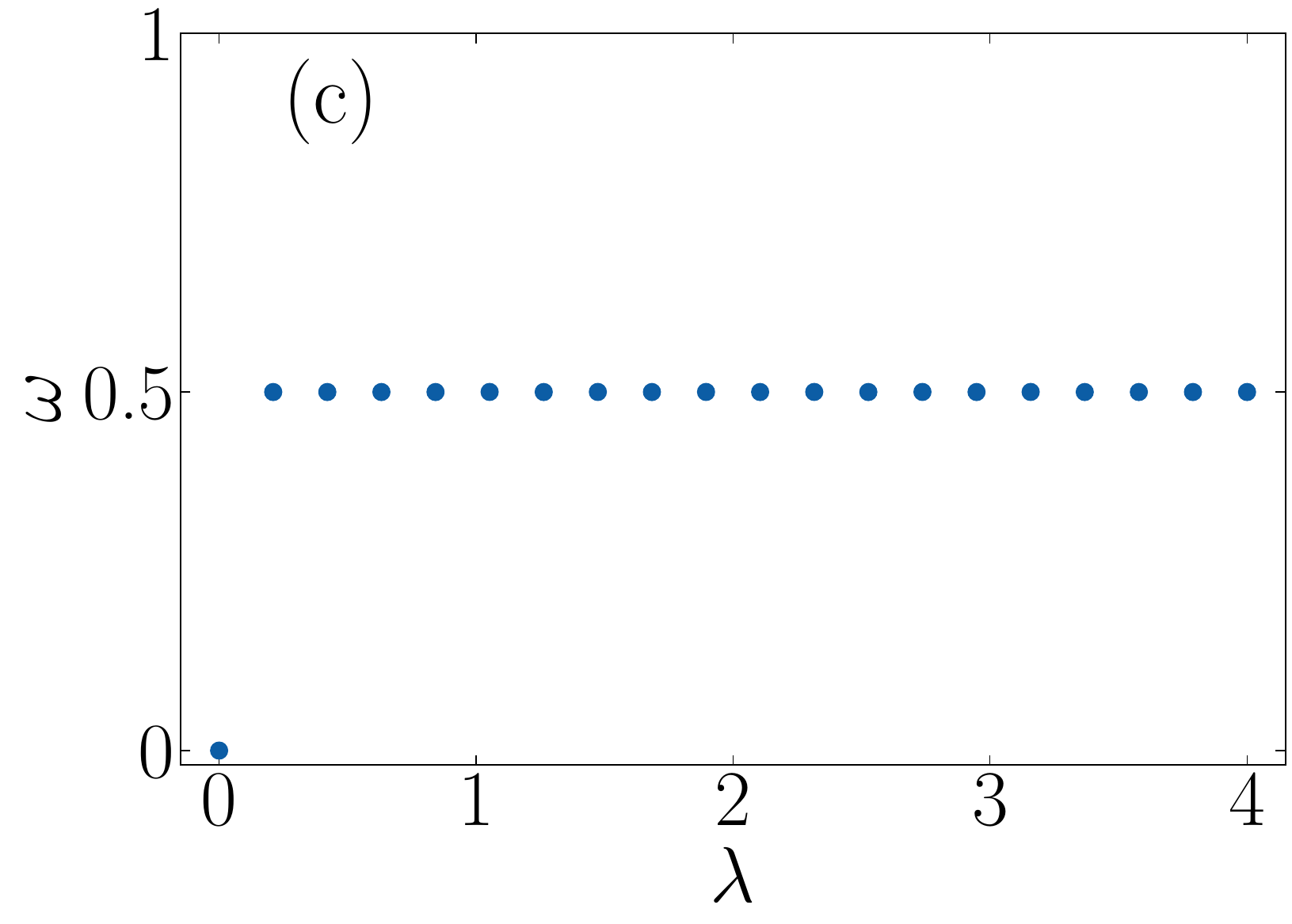}
	\includegraphics[width=42mm,height=28mm]{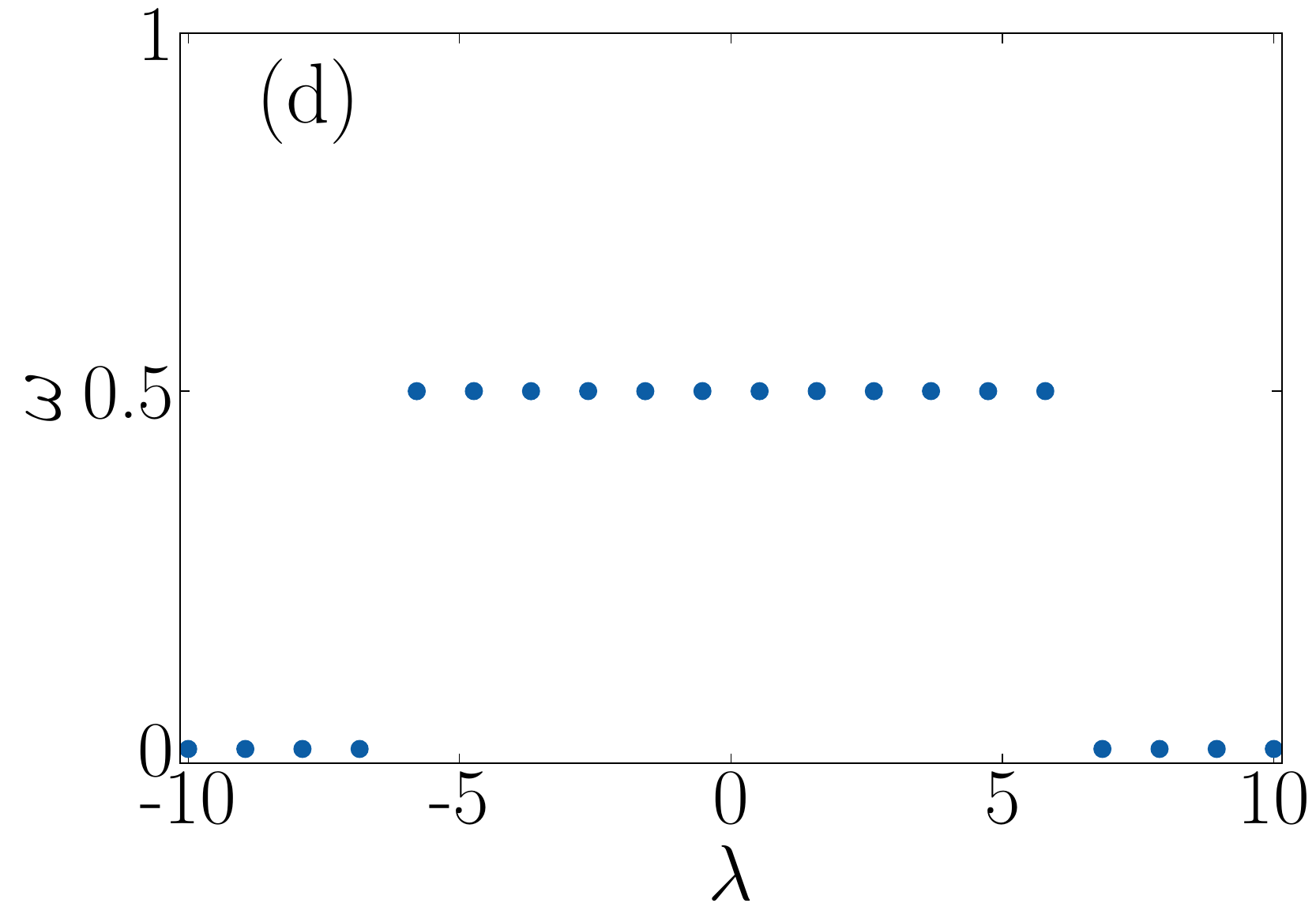}
	\caption{(a) and (b) show the complex spectrum for models I and II. The green dashed lines denote the AMEs. (c) and (d) show the winding number for model I and model II. Here the system size of models is $L=2584$, and the parameters $\lambda=2.0$ and $g=0.5$ under PBC.}
	\label{fig6}
\end{figure}

\section{Topological origin of non-Hermitian anomalous mobility edges}\label{sec:5}
The emergence of critical states and AMEs in the investigated models reveals a universal underlying mechanism. The underlying mechanism is rooted in the zeros of hopping coefficients in the thermodynamic limit or the presence of unbounded potentials within the Hamiltonian, which facilitate the existence of critical states. This mechanism applies not only to Hermitian systems but also to non-Hermitian systems, such as our models. In this section, we explore the real-complex spectrum transition and the topological origin of the AMEs in our two NH quasiperiodic models. Through numerical diagonalization of Hamiltonians (\ref{eq:model1}) and (\ref{eq:model2}) with specific parameters $\lambda=2.0$ and $g=0.5$ under PBC, we can obtain insights into this transition. The numerical results, depicted in Figs.~\ref{fig6} (a) and (b), indicate that the FD $\Gamma$ of real energies is nearly close to $0$, suggesting localization of the corresponding eigenstates. Conversely, for complex energies, the FD $\Gamma$ approaches $0.5$, indicating the corresponding eigenstates are critical. These findings suggest a localization-critical transition that co-occurs with the real-complex spectrum transition. A winding number can describe this topological transition. For the phase factor, $\phi$ of the potential in our NH models is continuously varied, the winding number can be defined as~\cite{longhi2019topological,zeng2020winding,liu2020generalized,liu2021exacta,longhi2021non1}
\begin{equation}\label{winding}
w(E_B)=\lim_{L \rightarrow \infty} \frac{1}{2 \pi i} \int_{\phi}^{2 \pi} d \phi \partial_{\phi}{\rm ln \; det} \left\{ H(\phi) -E_B \right\},
\end{equation}
which measures the change of the spectrum and topological transition for the base energy $E_B$ when $\phi$ is changed continuously from $0$ to $2\pi$. In Fig.~\ref{fig6} (c), we set the base energy in the middle of the energy spectrum $E_B=E_{mid}$, then the winding number $w=1/2$ when the AMEs emerge. Note that for fixed $g=0.5$ and except $\lambda=0$, the model I has AMEs for all quasiperiodic potential strengths $\lambda$, and thus, the system is always in topological AME phase coexisting with localized and critical states. However, for the numerical results of model II as shown in Fig.~\ref{fig6} (d), the winding number can change from $0$ to $1/2$ and then back to $0$ when changes the $\lambda=-10$ to $10$. This observation confirms a topological transition from a trivial localized phase to a topological AME phase with changing $\lambda$. Based on the above numerical results and discussions, we know that the emergence of such AMES in our models is topological, i.e., the energies of localized and critical states exhibit distinct topological structures in the complex energy plane. This is similar to NH topological ME separating localized and extended states in the complex energy plane as a result of NH terms in the quasicrystals.

\section{Conclusion}\label{sec:6}
In summary, we have studied the critical states and AMEs to 1D NH quasicrystals with nonreciprocal hopping. The study has observed two distinct mechanisms that lead to the emergence of robust critical states in the two NH models investigated. These robust critical states and AMEs are attributed to the zeros of hopping coefficients in the thermodynamic limit and the presence of unbounded quasiperiodic potentials. The AMEs and LEs can be analytically obtained from the NH proposed models using Avila's global theory. To confirm the emergence of robust critical states in both models, we perform a finite-size analysis of the MFD and level spacings of the eigenvalues. Furthermore, we demonstrate the localization-critical transition that co-occurs with the real-complex spectrum transition and the topological origin of the AMEs in our NH quasiperiodic models.

Our work contributes to developing critical states and AMEs for 1D NH quasicrystals. In future research, it may be valuable to extend the concept of AME to higher-dimensional systems~\cite{wang2023two,duncan2023critical} or other interacting systems~\cite{kohlert2019observation,hamazaki2019non,zhai2020many,wang2022non,li2023non,jiang2023stark,liu2023ergodicity,yu2024non,cheng2024stable}. Additionally, it would be intriguing to explore transport phenomena in NH quasiperiodic systems with critical states or AMEs.

\section*{Acknowledgments}
This work is supported by the China Postdoctoral Science Foundation (No.~2023M743267) and the National Natural Science Foundation of China (Grant No.~12304290, No.~12204432, and No.~62301505). LP also acknowledges support from the Fundamental Research Funds for the Central Universities.

\bibliography{Localization}
\end{document}